\documentclass[review,onefignum,onetabnum]{siamart190516}
\pdfoutput=1

\usepackage{amssymb}
\usepackage{amsfonts}
\usepackage{amsmath,mathtools}
\usepackage{algorithmic}

\author{Peijian Ding \thanks{This work was done while the author was an undergraduate student at Emory University. peijian.ding6@gmail.com}\and Advisor: Dr. James G. Nagy \thanks{Department of Mathematics, Emory University. jnagy@emory.edu}}
\title{Accelerated Alternating Minimization for X-ray Tomographic Reconstruction}
\date{July, 2021}
\begin{document}
\maketitle

\begin{abstract}
   While Computerized Tomography (CT) images can help detect disease such as Covid-19, regular CT machines are large and expensive. Cheaper and more portable machines suffer from errors in geometry acquisition that downgrades CT image quality. The errors in geometry can be represented with parameters in the mathematical model for image reconstruction. To obtain a good image, we formulate a nonlinear least squares problem that simultaneously reconstructs the image and corrects for errors in the geometry parameters. We develop an accelerated alternating minimization scheme to reconstruct the image and geometry parameters. 
\end{abstract}



\section{Introduction} \label{sec:intro}
Tomography is a technique of displaying representations of a cross section through an object through the use of some penetrating waves such as X-ray or ultrasound. In simple words, it allows us to see the inside of an object without breaking it. Thus, tomography is widely used in medical imaging, seismic exploration, and material science. In medical imaging, a Computerized Tomography (CT) Scan creates a cross-sectional image of human body by combining X-ray images taken from different angles.
\begin{figure}
    \centering
    \includegraphics[width=0.5\textwidth]{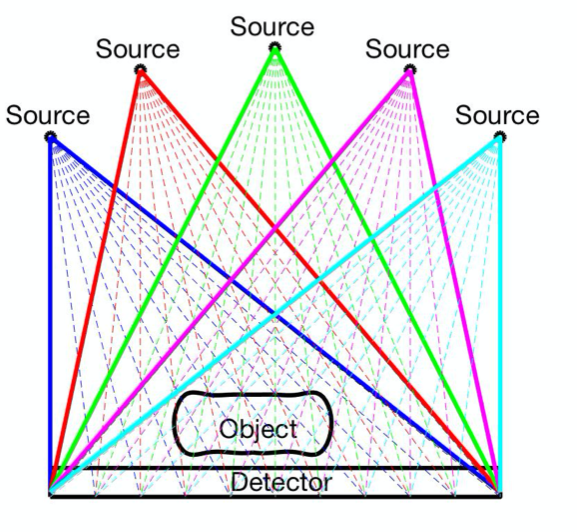}
    \caption{This is a simple illustration of our tomography problem.}
    \label{fig:2Dtomo}
\end{figure}

\begin{figure}
    \centering
    \includegraphics[width=0.33\textwidth]{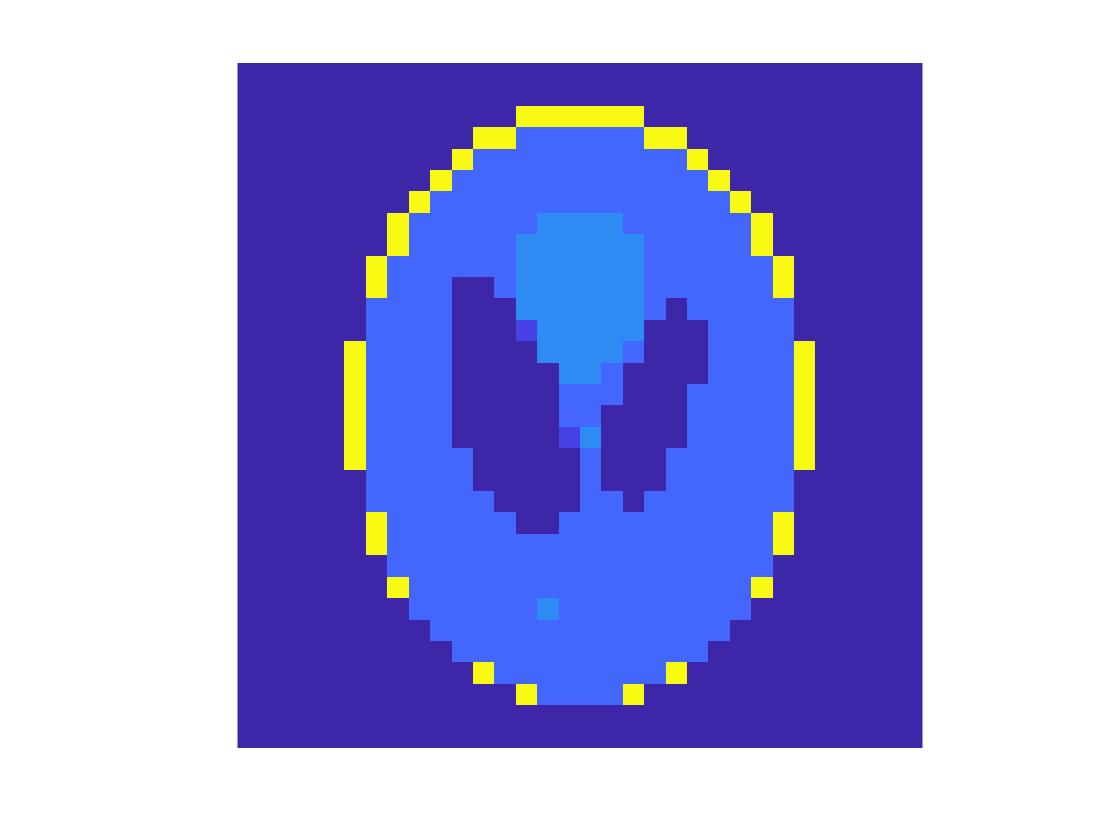}%
    \includegraphics[width=0.33\textwidth]{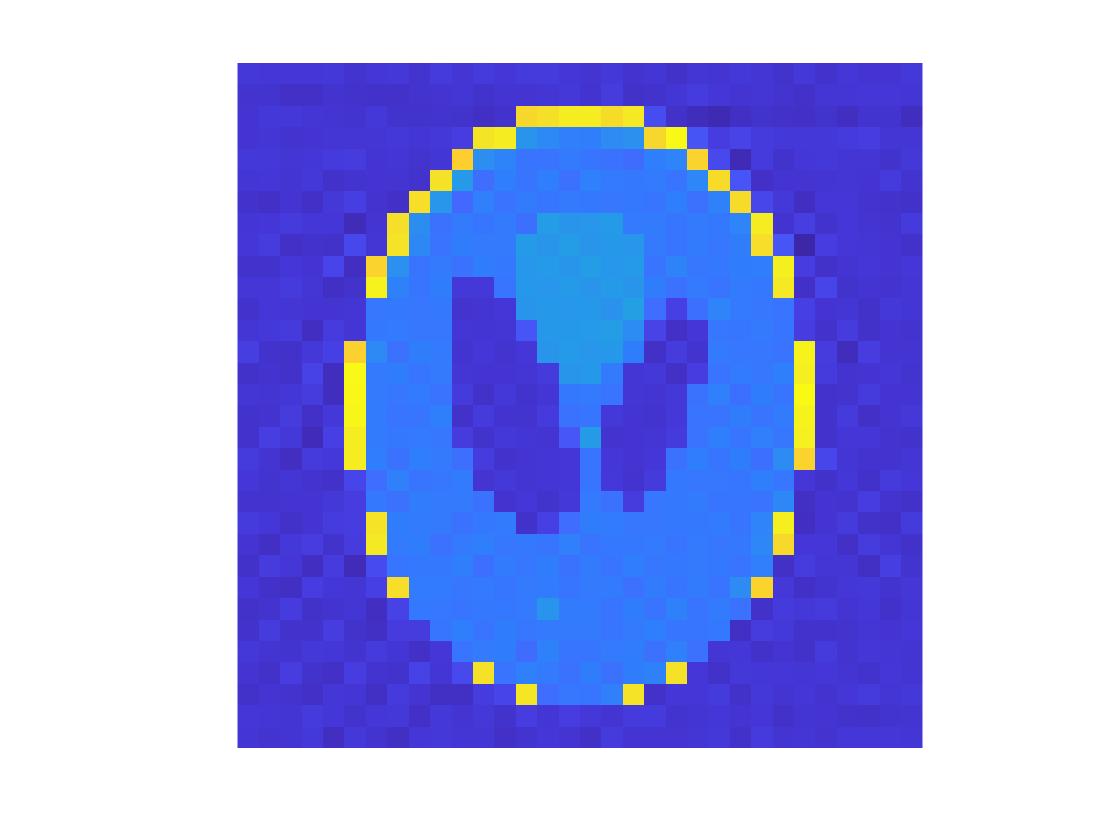}%
    \includegraphics[width=0.33\textwidth]{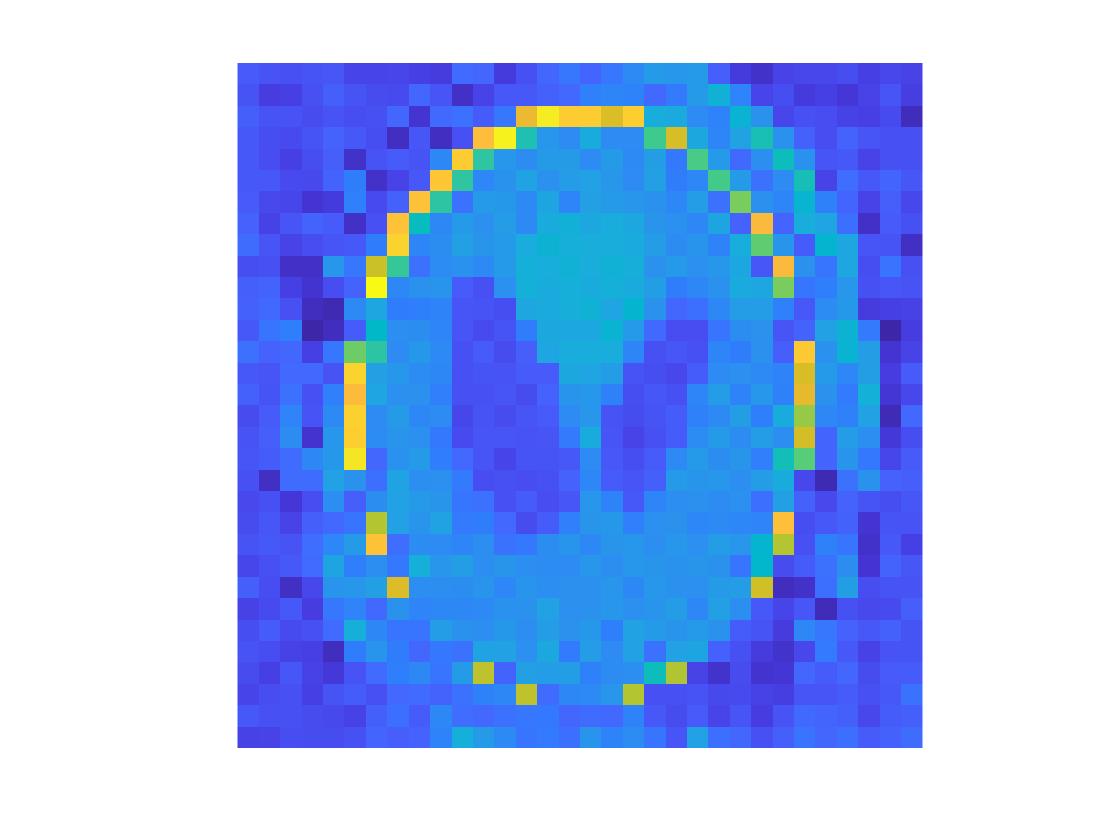}
    \caption{$32\times32$ true image of a Shepp–Logan phantom (left), image computed by taking into account the correct source-to-object distance (middle), image computed by using incorrect source-to-object distance (right)}
    \label{fig:motivation}
\end{figure}
During a CT scan, the patient lies on a bed that slowly moves through the gantry while the X-ray tube rotates around the patient and shoots X-ray beams through the human body, received by a detector. See Figure \ref{fig:2Dtomo}. Then, an image of the cross section of the human body is reconstructed following a mathematical procedure. \par

 Although CT images can help doctors diagnose and monitor diseases such as Covid-19, regular CT machines are heavy and expensive and not widely available in less developed areas. The goal of this paper is to compensate for cheaper and more portable machines by solving for geometry parameters such as the source-to-object distance that may not be calibrated precisely during the imaging process. \par 
Source-to-object distance measures how far away the center of the object is from the X-ray source. Since the source-to-object distance may vary from angle to angle, the reconstructed image will be corrupted if incorrect values are used, as illustrated in Figure \ref{fig:motivation}. Similar conclusions can be drawn if incorrect angles are used to reconstruct the image. Thus, our algorithm will significantly improve image quality by taking into account the variation in geometry parameters. \par

For the purpose of this paper, we primarily focus on 2D computed tomography to simplify notations and to reduce computational costs required to perform a substantial number of numerical experiments. But the methods discussed in this paper can be extended to $3$D problems. An X-ray imaging problem can be formulated as an inverse problem in linear algebra: 

$$Ax=b,$$
where $A$ is a $m\times n$ forward operator, $b$ is projection data, also referred to as a sinogram, and $x$ is the solution vector which represents the image. 

 Since the forward operator $A$ is determined by geometry parameters $r$, we form the following nonlinear least squares problem: 
\begin{equation}
     \underset{x,r}{\arg\min}||A(r)x-b||_2^2
     \label{eq:lsq_problem}
\end{equation}
where $x$ represents the image vector, and $A$ is the forward operator that is a function of $r$ and maps the true image $x$ to the sinogram $b$. In this paper, we denote $r$ as a general geometry parameter vector that can represent both source-to-object distance $d$ and errors in angles $\delta\theta$. Thus, the conclusions we make about $r$ also apply to both $d$ and $\delta\theta$. In this paper, we show that Equation \ref{eq:lsq_problem} can be solved using an accelerated Block Coordinate Descent method. In addition to incorporating an acceleration scheme, we also exploit separability to further reduce the computational cost.

\section{Alternating Minimization Scheme}\label{sec:AM}
In order to solve the nonlinear least squares problem, an intuitive approach to consider is the Block Coordinate Descent (BCD) algorithm. In this section, we discuss the linear and nonlinear least squares problem involved in BCD. We also briefly discuss the variable projection approach and argue the advantage of BCD over variable projection for our problem. 
\subsection{Block Coordinate Descent} \label{sec:separability}
Block Coordinate Descent is a simple approach to solve an optimization problem. Its idea is based on the general Coordinate Descent (CD) algorithm. Because of its lack of sophistication, most optimization researchers have not focused on this approach until recently when CD approaches were found to be computationally competitive to other reputable alternatives in various applications such as machine learning \cite{chap1_1}. Since we are able to exploit separability in the geometry parameters, the BCD method, which is given in Algorithm \ref{alg:AM},  is worth investigating for tomographic reconstruction.

\begin{algorithm}
\begin{algorithmic}[1]
\STATE{Input: $r_0 \in \mathbb{R}^{N_A}$, $x_0 \in \mathbb{R}^{n}$}
\FOR{$k=1,2,\dots$ until a stopping criterion holds}
\STATE{$r_{k} =\underset{r}{\arg\min} ||A(r)x_{k-1}-b||_2^2$}
\STATE{$x_{k} =\underset{x}{\arg\min} ||A(r_{k})x-b||_2^2$}
\ENDFOR
\end{algorithmic}
\caption{BCD to Reconstruct Geometry and Image Parameters}\label{alg:AM}
\end{algorithm}
\noindent $N_A$ denotes the number of angles, and $n$ is the length of the image vector.
Note that in practice an initial estimate $r_0$ is given. With this information, we can easily obtain $x_0$ by solving the linear least squares problem, $$x_0 = \underset{x}{\arg\min} ||A(r_{0})x-b||_2^2. $$ 

We remark that for ill-posed problems, computing $x_0$ and Step $4$ of Algorithm \ref{alg:AM} typically requires incorporating regularization procedures to avoid amplifying noise when solving the linear least squares problems. This is discussed further in section \ref{sec:regu}. There is another property in our matrix that makes this alternating minimization scheme favorable. The matrix $A$ and vector $b$ can be partitioned into block of rows as 
\begin{equation*}
    A = \begin{bmatrix}
    A_1(r^{(1)})\\
    \vdots \\
    A_i(r^{(i)})\\
    \vdots \\ 
    A_{N_A}(r^{(N_A)})
    \end{bmatrix} \quad
    b = \left [\begin{array}{cc}
         b_1  \\
         \vdots \\
         b_i \\
         \vdots \\
         b_{N_A}
    \end{array} \right ]
\end{equation*}
\noindent where errors in geometry parameters associated with rows in each $[A_i(r^{(i)}), b_i]$ block are assumed constant. For example, if we collect projections at $0,1,2,\dots, 359$ degrees around the center of the object, and $N_A = 10$, then an error is introduced into the geometry parameters once every $36$ degrees. In reality, $N_A=360$, but if some of the errors are small enough to ignore, using a smaller $N_A$ can reduce the computational cost. The choice of $N_A$ is discussed further in Section \ref{sec:NA}. \par
The separability of the geometry parameters allows us to solve for a set of much smaller systems at Step $3$ in Algorithm \ref{alg:AM}, namely, $$r_{k}^{(i)} = \underset{r}{\arg\min}|| A_i(r)x_{k-1}-b_i||_2^2, \quad i = 1,2,\dots,N_A$$ and then $$r_{k} =\left [ \begin{array}{cc}
     r_{k}^{(1)} \\
     \vdots \\ 
     r_{k}^{(N_A)}
\end{array} \right ]. $$
Note that parameters in the vector $r_{k}$ are completely independent, so they can be updated simultaneously on a parallel computing architecture, which could dramatically lower the computing time. In fact, if we only consider source-to-object distance as geometry parameters, the dimension of the parameter $r^{(i)}$ is one. Comparing to this alternating minimization scheme, Variable Projection \cite{varpro}\cite{varpro2} cannot utilize this matrix property because its matrix multiplication process would take away the separability property. Thus, it makes this alternating minimization scheme more worthwhile to investigate. \par
Next, we discuss linear least squares solvers and nonlinear least squares solvers respectively.
\subsection{Linear Least Squares Problem}
In this subsection, we consider the linear least squares problem in Step $4$ of Algorithm \ref{alg:AM}, $\underset{x}{\min}||Ax-b||_2^2$, where we assume matrix $A$ is fixed, e.g. $A=A(r_k)$. 

\subsubsection{Regularization} \label{sec:regu}
Typically in inverse problems, the data we obtain is not the exact data. This is also the case in our X-ray tomography problem, even in the case when geometry parameters, and hence matrix $A$, are known exactly. Ideally, we want to solve $Ax_{true}=b_{true}$ but the linear system that we actually solve is:
\begin{equation}
    Ax= b =b_{true}+\eta
\end{equation} where $\eta$ is the noise in our measurement data, $b_{true}$ is the noise-free data, $A\in \mathbb{R}^{m\times n}$, and $b\in \mathbb{R}^m$. For simplicity, we assume $A$ is full rank and $m=n$. Our discussion also holds for more general cases when we compute a pseudo-inverse instead of an exact inverse. \par
By using the SVD of $A=U\Sigma V^T$ where $\sigma_i$ is the $i^{th}$ singular value and $u_i$, $v_i$ are the $i^{th}$ column of the left and right orthogonal matrix $U$ and $V$, we first recall that $Av_i=\sigma_i u_i$, and we note that we can find scalars $\alpha_i$ and $\eta_i$ such that 

\noindent\begin{minipage}{.4\linewidth}
\begin{equation*}
    x_{true} = \sum_{i=1}^n \alpha_i v_i
\end{equation*}
\end{minipage}%
and 
\begin{minipage}{.4\linewidth}
\begin{equation*}
    \eta= \sum_{i=1}^n \eta_i u_i
\end{equation*}
\end{minipage} \\
\\
Using these relations, we obtain 
\begin{equation}
\begin{split}
    b &= \sum_{i=1}^{n}(\alpha_i \sigma_i + \eta_i)u_i
\end{split}
\end{equation}
Then, notice that 
\begin{equation}
    \begin{split}
        A^{-1}u_i& = V\Sigma^{-1}U^T u_i \\
        &= \frac{1}{\sigma_i} v_i
    \end{split}
\end{equation}

Thus, 
\begin{equation}
    \begin{split}
        x &=A^{-1}b\\
        &= \sum_{i=1}^{n}\left (\alpha_i + \frac{\eta_i}{\sigma_i} \right ) v_i
    \end{split}
\end{equation}

From this result we observe that the noise in the data is magnified by the small singular values. Since $\sigma_1 \geqslant \sigma_2 \geqslant ... \geqslant \sigma_n \geqslant 0 $, larger indices correspond to smaller singular values. The computed solution is dominated by noise amplified by division of small singular values. Thus, we need regularization schemes to filter out this noise. In particular, a regularized solution can be written as: 
\begin{equation}
    x_{reg} = \sum_{i=0}^n \phi_i \left(  \alpha_i +\frac{\eta_i}{\sigma_i}\right) v_i
\end{equation}
where the scalar $0 \leqslant \phi_i \leqslant 1$ is called a filter factor. As $\sigma_i$ decreases, the filter factors should approach zero so that the noise contributed by the small singular values are filtered out.\par

\begin{figure}[ht]
     \centering
     \includegraphics[width=0.5\textwidth]{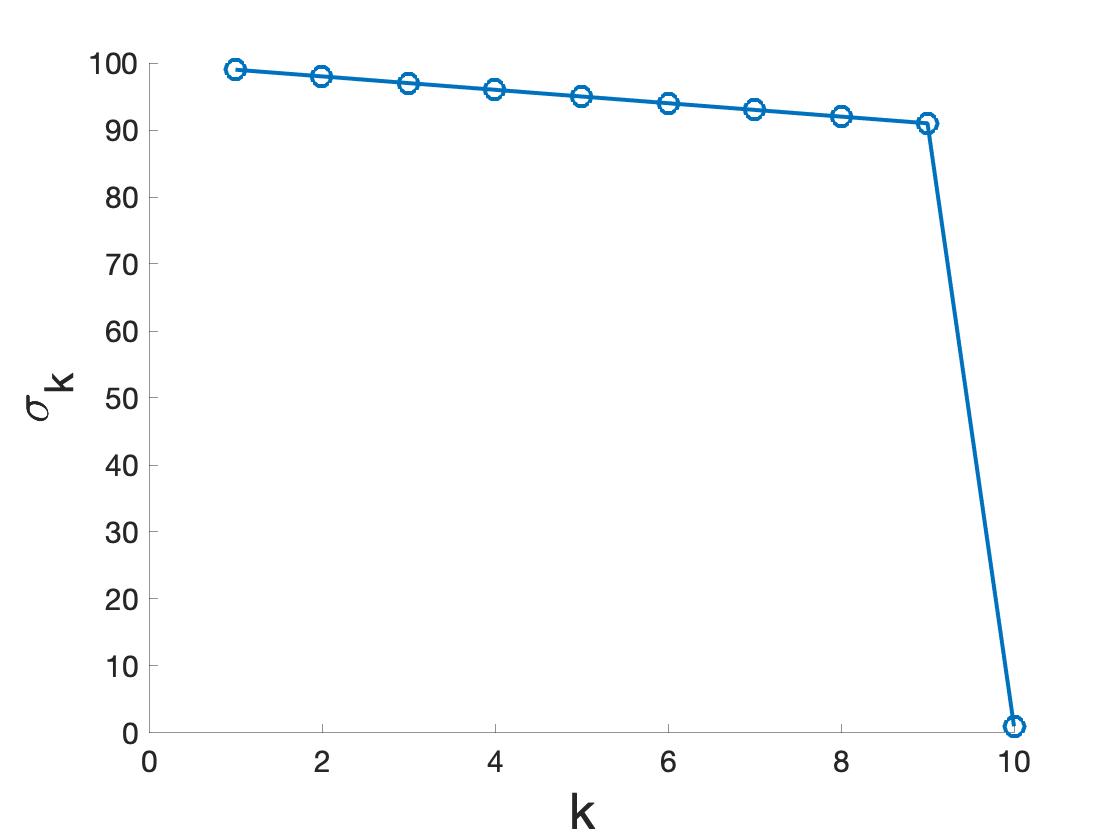}
     \caption{The singular values plot of a $10\times10$ diagonal matrix whose singular values are respectively $99$, $98$ ...  $91$, and $1$}
     \label{fig:TSVD}
 \end{figure}

\paragraph{Truncated Singular Value Decomposition} Since the noise is magnified by small singular values, the most intuitive approach is to cut off the small singular values by setting them to zero. This is called Truncated SVD regularization. The TSVD solution to the inverse problem is given by 
\begin{equation}
    x_{reg} = \sum_{i=1}^k \frac{b^Tu_i}{\sigma_i}v_i
\end{equation}
 where $k \leqslant n$. The critical part in the TSVD is identifying the threshold $k$. One approach is choosing $k$ at a significant drop-off of singular values, as illustrated by Figure \ref{fig:TSVD}.
  \par
\begin{figure}[ht]
     \centering
     \includegraphics[width=0.5\textwidth]{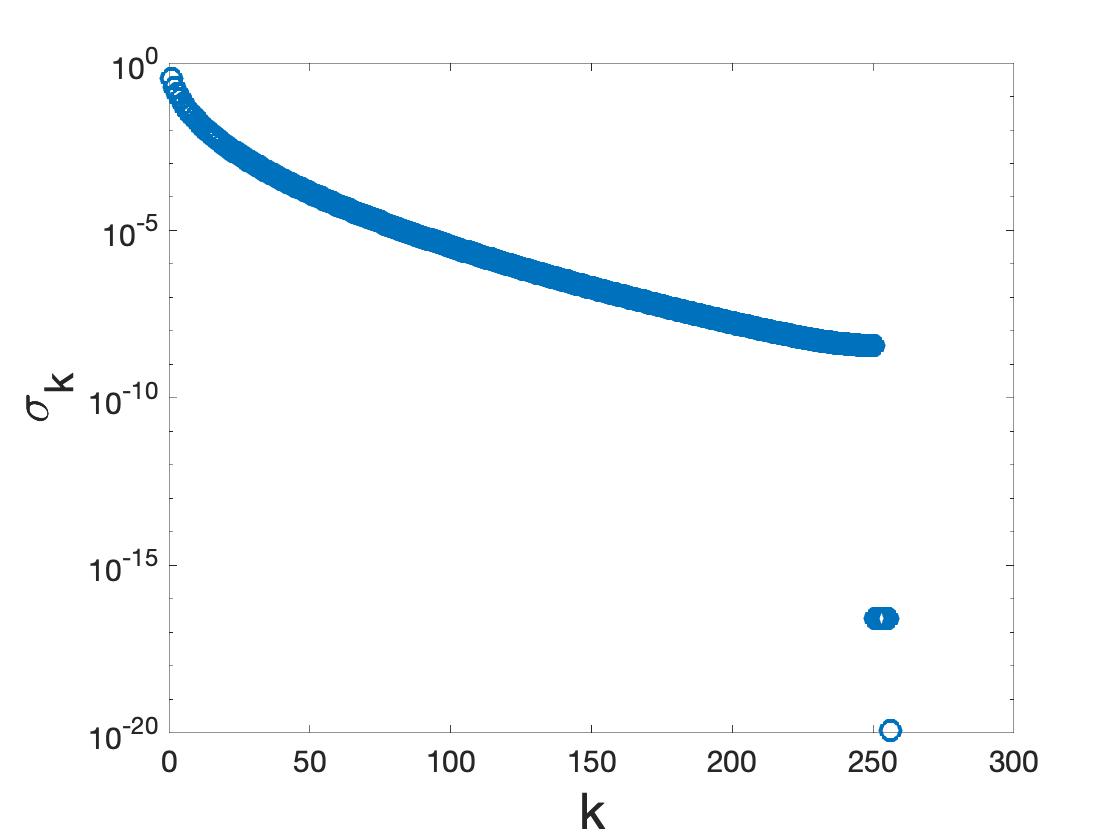}
     \caption{The singular values of the $256 \times 256$ test problem \emph{heat} generated from Regularization Tools in MATLAB}
     \label{fig:SmoothDecay}
 \end{figure}

 In this case, $k=9$ can be easily identified as the threshold for the TSVD approach. However, typically in inverse problems, singular values decay smoothly, as illustrated by Figure \ref{fig:SmoothDecay}.
 In cases like Figure \ref{fig:SmoothDecay}, a reliable cut-off threshold for singular values is hard to find. Note that $k=250$ is not a good choice for the cut-off because the singular values are very close to zero for smaller indices $k$ (e.g. $\sigma_{200} \approx 10^{-8}$). Therefore, we need more advanced regularization techniques to deal with smoothly decaying singular values. 
 
\paragraph{Tikhonov Regularization} Classical Tikhonov Regularization can be used to solve ill-posed problems. The regularized solution $x_{reg}$ is the unique solution to the following: 
 \begin{equation}
  \label{eq:tik}
     \underset{x}{\min} {||Ax-b||_2^2 + \lambda^2||x||_2^2}
 \end{equation}
 where $\lambda$ is the regularization parameter that controls the smoothness of the regularized solution. The above equation is equivalent to the following: 
 \begin{equation}
     \underset{x}{\min} \left\| \begin{bmatrix}
     A \\
     \lambda I
     \end{bmatrix} x - 
     \begin{bmatrix}
     b\\
     0
     \end{bmatrix}\right \| _2^2
 \end{equation}
Then the normal equations for this least squares problem can be written as: 
\begin{equation}
    (A^TA + \lambda^2I)x_{reg} = A^Tb.
\end{equation}
From the normal equations, we obtain the following: 
\begin{equation}
\label{eq:reg-tik}
    \begin{split}
        x_{reg} & = (A^TA+\lambda^2I)^{-1}A^Tb \\
        &= \sum_{i=1}^{n} \phi_i \frac{b^Tu_i}{\sigma_i}v_i
    \end{split}
\end{equation}
where the filter factor is $\phi_i= \frac{\sigma_i^2}{\sigma_i^2+\lambda^2}$. For simplicity, we assumed $A$ has full rank. The pseudo-inverse can be used for the rank deficient case and Equation \ref{eq:reg-tik} would sum from $i=1$ to $rank(A)$ instead of the full column rank $n$. The modified Tikhonov regularization is not too much different from the classical Tikhonov except that it uses the 2-norm of $Lx$ instead of $x$ in Equation \ref{eq:tik}, where $L$ is a $p\times n$ matrix with $p
\leqslant n$. 
\begin{equation*}
    \underset{x}{\min} {||Ax-b||_2^2 + \lambda^2||Lx||_2^2}
\end{equation*}\par
In contrast to TSVD where singular values after $\sigma_k$ are cut off, the Tikhonov regularization applies a smoother filter to all the singular values. Given a good regularization parameter, the large singular values would not be affected too much whereas the small values would be gradually filtered more as they approach zero. The quality of the filtering depends on how we choose the regularization parameter. If $\lambda =0$, then all $\phi_i =1$ and we are directly calculating the inverse (or pseudo-inverse) solution. If we select a very large $\lambda \gg \sigma_1$, all $\phi_i$ would approach zero and we would have over-smoothed the solution.

\subsubsection{Parameter Choice Methods}
\label{sec:gcv}
The choice of regularization parameter is critical to the quality of the regularized solution. Parameter choice methods can usually be divided into two classes depending on their assumption about error norm $||\eta||_2^2 = ||b-b_{true}||_2^2$ where $b$ is the measured data and $b_{true}$ is the noise free data \cite{Hanson1}. The first class contains methods based on a good estimate of  $||\eta||_2^2$. The second class includes methods that are not based on a good estimate of  $||\eta||_2^2$, but seek to extract this information from the given right hand side $b$. In this section, we introduce the Generalized Cross-Validation method (GCV) which is a popular method in the second class. The underlying idea of GCV is that a good regularization parameter should predict missing values. For example, if some data point $b_i$ is missing in the right hand side, the regularized solution should predict this missing value well. GCV can be written as: 

\begin{equation}
   G (\lambda) = \frac{n||Ax_{reg}-b||_2^2}{trace(I_m -AA^\dagger_F)^2}
\end{equation}
where $x_{reg}$ is the regularized solution, $A^\dagger_F = \sum_{i=1}^n \phi_i\frac{ v_i u_i^T}{\sigma_i} = V\Sigma^\dagger_F U^T$, and $\Sigma^\dagger_F = diag(\frac{\phi_1}{\sigma_1}, \frac{\phi_2}{\sigma_2}, ..., \frac{\phi_n}{\sigma_n})$. 
The goal is to find $\lambda$ such that $G(\lambda)$ is minimized. Now we simplify the numerator and the denominator of $G(\lambda)$ respectively: 

\begin{equation*}
    \begin{split}
    ||Ax_{reg}-b||_2^2 &=  ||\Sigma V^T x_{reg} - U^Tb||_2^2 \\
    &= ||(\Sigma \Sigma^\dagger_F  - I)U^Tb||_2^2
    \end{split}
\end{equation*}
 
 \begin{equation*}
 \begin{split}
          trace(I_m -AA^\dagger_F) &=  trace((I_m -U\Sigma\Sigma_F^\dagger U^T))\\
           & =trace(I_m -\Sigma\Sigma_F^\dagger)
 \end{split}
 \end{equation*}
Thus, $G(\lambda)$ becomes:
\begin{equation}
\label{eq:gcv}
    G(\lambda) = \frac{n||(\Sigma \Sigma^\dagger_F  - I)U^Tb||_2^2}{trace(I_m -\Sigma\Sigma_F^\dagger)}
\end{equation}
Since $\phi_i = \frac{\sigma_i^2}{\sigma_i^2 + \lambda^2}$ in the Tikhonov case, 
\begin{equation*}
    \Sigma_F^\dagger = diag\left (\frac{\phi_1}{\sigma_1}, \frac{\phi_2}{\sigma_2}, ..., \frac{\phi_n}{\sigma_n} \right ) = diag \left (\frac{\sigma_1}{\sigma_1^2 + \lambda^2},\frac{\sigma_2}{\sigma_2^2 + \lambda^2},..., \frac{\sigma_n}{\sigma_n^2 + \lambda^2}\right) 
\end{equation*} 
Then we can write $G(\lambda)$ in the case of Tikhonov regularization as: 
\begin{equation}
    G(\lambda) = \frac{n\sum_{i=1}^n (\frac{\hat{b_i}}{\sigma_i^2 +\lambda^2})^2}{(\sum_{i=1}^n\frac{1}{\sigma_i^2+\lambda^2})^2}
\end{equation}
where $\hat{b} = U^T b$. We can solve for $\lambda$ by using the function \textbf{fminbnd} in MATLAB which is based on golden section search and parabolic interpolation. In practical applications, several studies have found that occasionally GCV would drastically under-smooth the solution by choosing the regularization parameter too small \cite{GCVunder} \cite{GCVunder2} \cite{GCVunder3}. In \cite{IRhybrid}, GCV has been found to over-smooth the solution in the Lanczos-hybrid methods, discussed in the next subsection. To alleviate this difficulty, it was proposed to use the weighted GCV method:
\begin{equation}
    G(w,\lambda) = \frac{n||Ax_{reg}-b||_2^2}{trace(I_m -wAA^\dagger_F)^2}
\end{equation}
where $w$ is the weight parameter that determines the function $G(w,\lambda)$ along with $\lambda$. When $w=1$, we have the non-weighted version of GCV like in Equation \ref{eq:gcv}. When $w>1$, the solution is smoother. When $w<1$, the solution is less smooth. So far in weighted-GCV literature, only experimental approaches have been used to determine the value for $w$.

\subsubsection{Hybrid LSQR}
\label{sec:hybrid}
We have so far discussed regularization and parameter choice methods using the SVD. For large-scale problems, such as in image reconstruction, directly applying SVD based methods is not computationally feasible. In this subsection we describe a hybrid LSQR scheme that combines an efficient iterative method with SVD based approaches that enforce regularization on small projected sub-problems. \par
The standard LSQR algorithm projects the linear least squares problem onto a sequence of Krylov subspace of small and increasing dimensions. The best approximation $x_k$ to the least squares problem in the Krylov subspace $K_k$ is given by $x_k = V_k y_k$, where $V_k$ is a $n \times k $ matrix with orthonormal columns at the $k^{th}$ step of the Golub-Kahan bidiagonalization process \cite{LSQR}. $y_k$ can be solved by 
\begin{equation*}
    y_k = \arg\min_{y_k} ||B_k y_k - \beta_1 e_1||_2
\end{equation*} 
where $B_k$ is the bidiagonal matrix at the $k^{th}$ step of the Golub-Kahan bidiagonalization process, $\beta$ is a scalar, and $e_1$ is the standard basis vector. When being applied to ill-posed problems, LSQR exhibits a semi-convergence behavior which means that early iterations construct information related to the solution while later iterations construct information related to noise \cite{IRhybrid}.\par This can be compensated by applying a direct regularization method such as Tikhonov or TSVD, which can be solved cheaply on a small scale problem of the reduced linear least squares in the Krylov subspace. So, we can write the Hybrid LSQR using Tikhonov regularization as: 
\begin{equation*}
    \underset{y_k}{\min} ||B_k y_k - \beta_1 e_1||_2^2 + \lambda_k^2 ||y_k||_2^2
\end{equation*}
where $\lambda_k$ is a regularization parameter chosen at the $k^{th}$ iteration using the weighted GCV method discussed in Section \ref{sec:gcv}. A method like GCV can be used to choose a stopping iteration so that $k$ will not be too large; details can be found in \cite{IRhybrid}.

Comparing to LSQR, this hybrid method can effectively stabilize the iterations \cite{IRhybrid}. Although at each iteration a new regularization parameter must be chosen, it is not computationally expensive for the projected problem. \par
To summarize the method, the hybrid LSQR method projects the large scale linear least squares problem onto a low-dimensional Krylov subspace where we can inexpensively apply a direct regularization method like the adaptive weighted-GCV. \par

\subsection{Nonlinear Least Squares}

In our alternating minimization scheme, we iteratively solve the image and the geometry parameters. While we have discussed methods to solve the linear least squares problem in the previous section, we need other tools to solve the nonlinear least squares problem:
\begin{equation}
    \underset{r}{\min} ||A(r)x-b||_2
\end{equation}
where $x$ is approximated by the linear least squares solution we obtained by using hybrid LSQR in Section 2.3. \par
We utilize the implicit filtering method which solves the bound-constraint optimization problem for which the derivative information is not available \cite{CTKelley}. Since we do not have the derivative information of our objective function and a reasonable bound can be established for the geometry parameters in our tomographic reconstruction problem, implicit filtering serves as a good tool to solve our problem. Implicit filtering builds the local model of the objective function using a quasi-Newton method. \par
In our numerical experiments, we compare implicit filtering to the MATLAB function \textbf{fminbnd}.\par

\section{Acceleration Algorithms}
\label{sec:acc}

In the previous section, we have introduced the alternating minimization scheme and the methods we use to solve least squares problems. In this section, we introduce  methods that will accelerate the convergence of our minimization scheme. 

\subsection{Accelerated Block Coordinate Descent}
Since we can divide variables in our least squares problem into two blocks – geometry parameters $r$ and image $x$, it makes sense for us to directly investigate methods that accelerate the BCD algorithm. We implemented \textbf{Accelerated Block Coordinate Descent} (ABCD). This method can be applied to a four-block problem by dividing it into two larger blocks, but in our problem we do not have to do so. Theoretically, the proposed acceleration method in \cite{ABCD} has a complexity of $O(\frac{1}{k^2})$. In our implementation, we simplify the algorithm as Algorithm \ref{alg:imABCD}.
\begin{algorithm}[H]
\caption{Accelerated Block Coordinate Descent}\label{alg:imABCD}
\begin{algorithmic}[1]
\STATE{Inputs: $t_0 =1$, $r_0 \in \mathbb{R}^{N_A}$ and $x_0 \in \mathbb{R}^n$ }

\FOR{$k=1, 2, \dots$ until a stopping criterion holds}
\STATE $\Tilde{r}_{k} =\underset{r}{\arg\min} ||A(r)x_{k-1}-b||_2^2$

        $\Tilde{x}_{k} = \underset{x}{\arg\min} ||A(\Tilde{r}_{k})x-b||_2^2 + \lambda^2 ||x||_2^2$ \;
        
\STATE $\Tilde{w}_k = (\Tilde{x}_{k},\Tilde{r}_{k})$ 

        $t_{k} = \frac{1}{2} (1+\sqrt{1+4t^2_{k-1}})$ 
        
        $w_{k} = \Tilde{w}_{k-1}+\frac{t_{k-1}}{t_{k+1}}(\Tilde{w}_{k}-\Tilde{w}_{k-1})$, where $w_k = (x_k, r_k)$
\ENDFOR
\end{algorithmic}
\end{algorithm} 
\noindent $N_A$ denotes the number of angles, and $n$ is the length of the image vector. We remark that we can exploit separability when solving for $\Tilde{r}_k$, as discussed in Section \ref{sec:separability}. As mentioned in Algorithm \ref{alg:AM}, since $r_0$ is given, $$x_0 = \underset{x}{\arg\min} ||A(r_{0})x-b||_2^2 + \lambda^2 ||x||_2^2. $$  \par

The Tikhonov regularized least squares problems for $x_0$ and $\Tilde{x}_k$ are solved using the hybrid scheme \textbf{IRhybrid\_lsqr} provided in \textbf{IRTools}\cite{IRtools} in MATLAB. The nonlinear least squares problem for $\Tilde{r_k}$ is solved using a MATLAB package called \textbf{imfil} \cite{CTKelley}. We have found that accelerating the solution vector $x$ alone has yielded stabler results with slightly better accuracy than performing acceleration on both $x$ and $r$. We show the numerical experiments in section \ref{sec:experiments}. 

\subsection{Anderson Acceleration}
Anderson Acceleration, also called Anderson mixing, is a method used to accelerate the convergence of fixed point iteration. Note that we can write out alternating minimization scheme as a fixed point iteration,
\begin{algorithm}[h]
    \caption{Fixed point iteration of the image vector} \label{alg:fixedpoint}
    \begin{algorithmic}[1]
    \STATE{Input: $x_k \in \mathbb{R}^n$, Output: $x_{k+1}=g(x_k)$}
    \STATE $r_{k+1} =\underset{r}{\arg\min} ||A(r)x_{k}-b||_2^2$
    
    \STATE $x_{k+1} = \underset{x}{\arg\min} ||A(r_{k+1})x-b||_2^2 + \lambda^2 ||x||_2^2$
\end{algorithmic}
\end{algorithm}
\noindent where $g:\mathbb{R}^n \rightarrow \mathbb{R}^n$ is the fixed point iteration of image vector $x$, as shown in Algorithm \ref{alg:fixedpoint}.\par
For this fixed point iteration, the general form of Anderson Acceleration is formed as the following:\par
\begin{algorithm}[h]
\caption{Anderson Acceleration}
\label{alg:AA}
\begin{algorithmic}[1]
\STATE{Inputs: $x_0$ and $m \geqslant 1$}
\STATE Set $x_1 = g(x_0)$, using Algorithm \ref{alg:fixedpoint}
\FOR{$k=1, 2, \dots$ until a stopping criterion holds}
\STATE $m_k = min(m,k$)
\STATE Set $F_k = (f_{k-m_k},...,f_k)$, where $f_i=g_i(x_i) - x_i$ and $g_i(x_i)$ comes from Algorithm \ref{alg:fixedpoint}
\STATE Determine $\alpha^{(k)} = (\alpha_0^{(k)},...,\alpha_{m_k}^{(k)})^T$ that solves

$\underset{\alpha}{\min} ||F_k\alpha||_2$ s.t $\sum_{i=0}^{m_k} \alpha_i = 1$
\STATE Set $x_{k+1} = \sum_{i=0}^{m_k} \alpha_i^{(k)}g(x_{k-m_k+i})$,
where $g(x_{k-m_k+i})$ is from Algorithm \ref{alg:fixedpoint}
\ENDFOR
\end{algorithmic}
\end{algorithm}
We can cast the linear constrained optimization problem in Step $7$ of Algorithm \ref{alg:AA} into an unconstrained form which is straightforward to solve and convenient for efficient implementation \cite{walker}.\par
We define $\nabla f_i = f_{i+1} - f_i$ for each $i$ and set $\nabla F = (\nabla f_{k-m_k},...,\nabla f_{k-1})$. Then the least squares problem is equivalent to 
$$\underset{\gamma= (\gamma_0,...\gamma_{m_k-1})^T}{min} ||f_k - F_k \gamma||_2,$$
where $\alpha_0=\gamma_0$ and $\alpha_i = \gamma_i - \gamma_{i-1}$, for $1 \leqslant i \leqslant m_k-1 $ and $\alpha_{m_k}= 1-\gamma_{m_k-1}$.\par
This unconstrained least squares problem leads to a modified version of Anderson Acceleration in Algorithm \ref{alg:AAunconstrained}, 

 \begin{algorithm}[ht]
    \caption{Modified Anderson Acceleration} \label{alg:AAunconstrained}
    \begin{algorithmic}[1]
    \STATE Given $x_0$ and $m \geqslant 1$ 
    \STATE Set $x_1 = g(x_0)$, using Algorithm \ref{alg:fixedpoint}
\FOR{k=1,2,...} 
\STATE $m_k = min(m,k$) 
\STATE Determine $\gamma^{(k)} = (\gamma_0^{(k)},...,\gamma_{m_k-1}^{(k)})^T$ that solves 

$\underset{\gamma= (\gamma_0,...\gamma_{m_k-1})^T}{min} ||f_k - F_k \gamma||_2$ 
\STATE Set $x_{k+1} = g(x_k)-G_k\gamma^{(k)}$, where $g(x_k)$ comes from Algorithm \ref{alg:fixedpoint}
\ENDFOR
\end{algorithmic}
\end{algorithm}
\noindent where $$x_{k+1} = g(x_k)- \sum_{i=1}^{m_k-1}\gamma_i^{(k)}[g(x_{k-m_k+i+1}) - g(x_{k-m_k+i})] = g(x_k)-G_k\gamma^{(k)} $$ with $G = (\nabla g_{k-m_k},...,\nabla g_{k-1})$, $\nabla g_i = g(x_{i+1})-g(x_i)$.\par
Homer Walker proposed implementation that efficiently updates the QR factors in the decomposition $F_k = Q_k R_k$ \cite{walker}. The basic logic is the following: every $F_k$ is obtained from $F_{k-1}$ with a column added on the right. If the resulting matrix has more columns than $m$, then delete one from the left. The column addition can be achieved by a modified Gram–Schmidt process. The deletion process is a little more complicated. We delete the first column on the left when $m_{k-1} = m$. If $F_{k-1}=QR$, then $F_{k-1}(:,2:m) = QR(:,2:m)$, where $R(:,2:m)$ is upper-Hessenberg. Then, we can determine $m$ Givens rotations to cancel out the entries in the sub-diagonal.

\section{Numerical Experiments}
\label{sec:experiments}
In this section, we make a few comparisons of different methods to solve the X-ray  tomography problem. Firstly, we compare the speed of BCD exploiting the separability of geometry and the speed of BCD without using such property. Secondly, we compare results produced from different number of angles. Thirdly, we make comparisons of all the acceleration schemes. Fourthly, different regularization parameters in the linear least squares solvers are compared. Fifthly, we compare implicit filtering with the MATLAB function \textbf{fminbnd} as the nonlinear least squares solver. Lastly, we show our algorithm works for both geometry parameters $d$ and $\delta\theta$. For experiments 1-5, we solve problems with only unknown source-to-object distance $d$, where geometry parameters $r = d$. For the last experiment, $r= (d, \delta\theta)$. Comparisons in all experiments are made about geometry errors and reconstruction errors. Geometry errors are defined as the relative errors of geometry parameters, $\frac{||r-r_{true}||_2}{||r_{true}||_2}$. For experiment 1-5, where $r=d$, geometry errors are $\frac{||d-d_{true}||_2}{||d_{true}||_2}$. For experiment 6, where $r=(d, \delta\theta)$ geometry errors are represented by both $\frac{||d-d_{true}||_2}{||d_{true}||_2}$ and $\frac{||\delta\theta-\delta\theta_{true}||_2}{||\delta\theta_{true}||_2}$. The reconstruction errors are defined as the relative errors of the image, $\frac{||x-x_{true}||_2}{||x_{true}||_2}$. \par
We use fan-beam projection for all our tomography problems for the sake of consistency. Note that we can also easily adapt our code to solve parallel beam projection problems by using the \textbf{IRtools} \cite{IRtools} and \textbf{AIR Tools} \cite{AIRtools} MATLAB packages. \par
In practice, good initial guesses of geometry parameters $r$ are available and prior knowledge can help us set the bounds for them. For the source-to-object distance $d$, we generate a test problem where true $d$ values, $d_{true}$, are random numbers (chosen from a uniform distribution) between $1.5$ and $2.5$. We use a constant initial guess of $d=2$ for all angles and set the bounds for $d$ from $1.5$ to $2.5$. For the errors in angles $\delta\theta$, we assume for each set of angles the error bound is from $-0.5$ to $0.5$. The test image we use is Shepp-Logan Phantom \cite{Shepp}, with image size $n = 32 \times 32$. The noise level is $ \frac{||\eta||_2}{||b||_2} = 0.01$, where $\eta$ is a vector with random entries chosen from a normal distribution. Budget is a hyper-parameter in \textbf{imfil} that stands for the maximum number of function evaluations in the nonlinear least squares solver. $N_A$ stands for number of angles. Moreover, $0^{th}$ iteration is included in the relative errors figures below. This represents the relative error of the initial guesses with regard to the true solution.\par

\subsection{BCD Exploiting Separability vs BCD}
\begin{figure}[ht]
    \centering
    \includegraphics[width=0.5\textwidth]{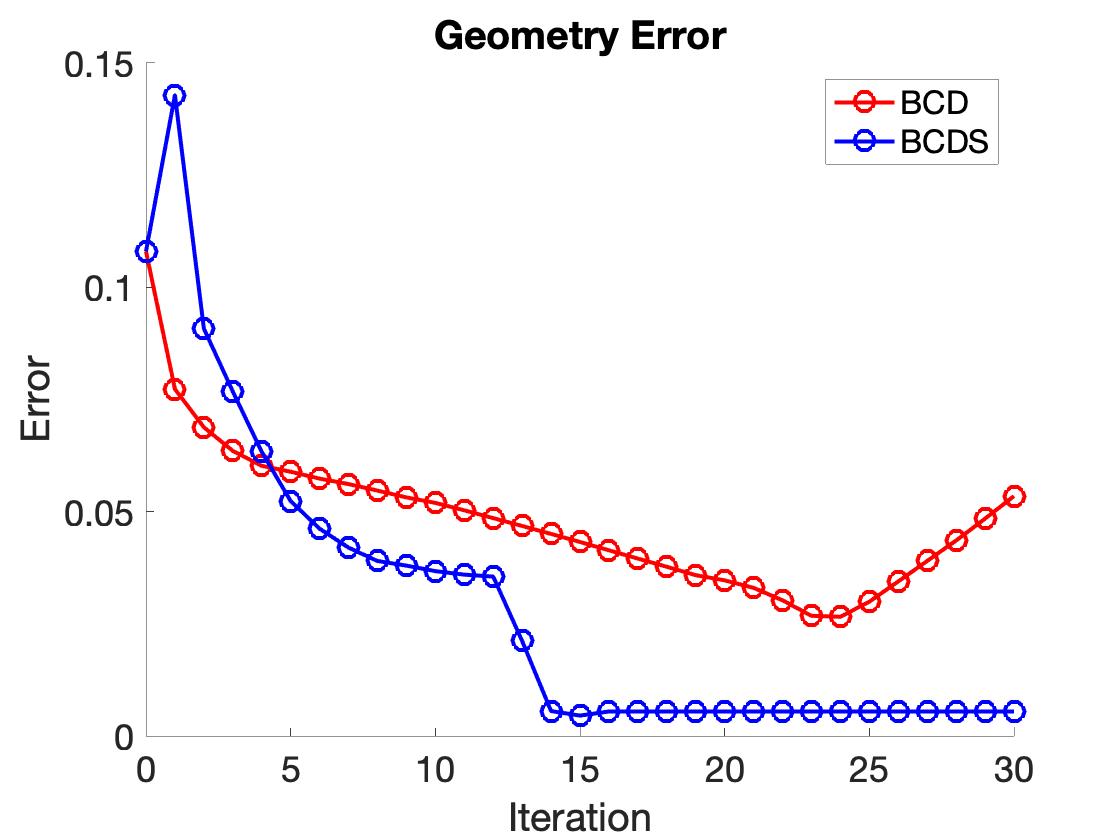}%
    \includegraphics[width=0.5\textwidth]{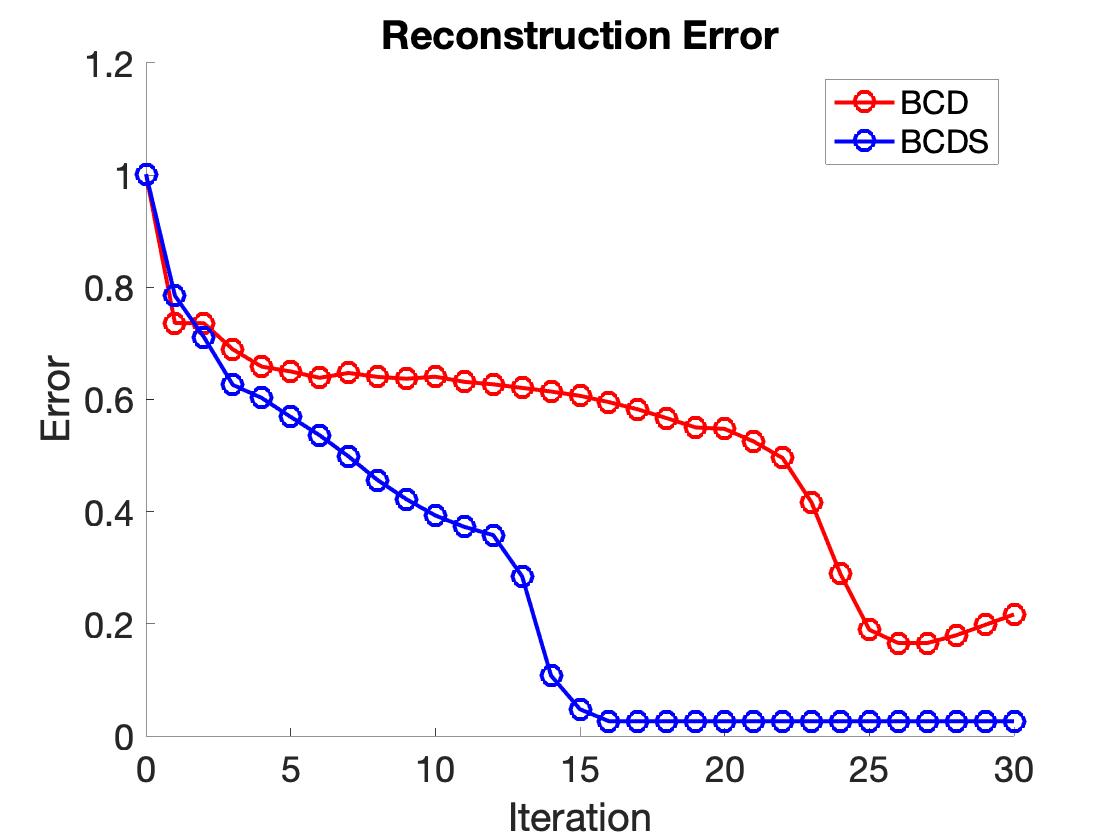}
    \caption{Comparison between BCD and BCDS.}
    \label{fig:BCDvsBCDS_R}
\end{figure}

\begin{figure}[ht]
    \centering
    \includegraphics[width=0.33\textwidth]{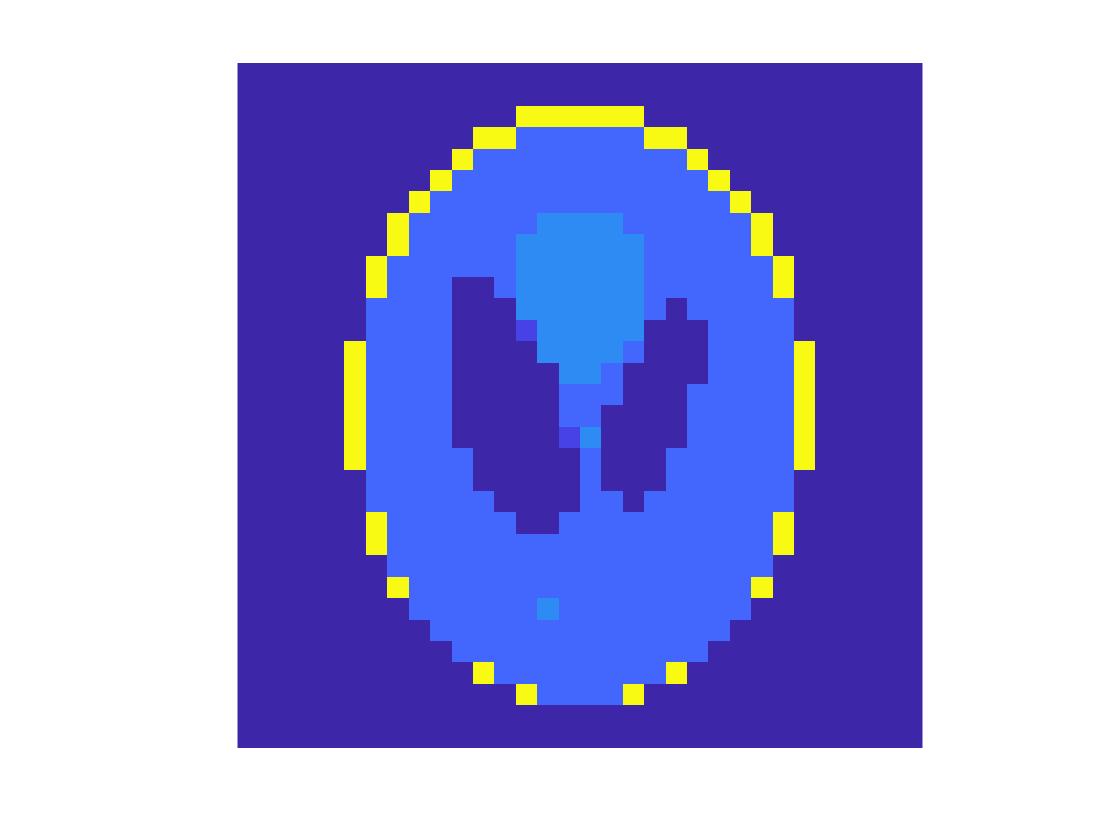}%
    \includegraphics[width=0.33\textwidth]{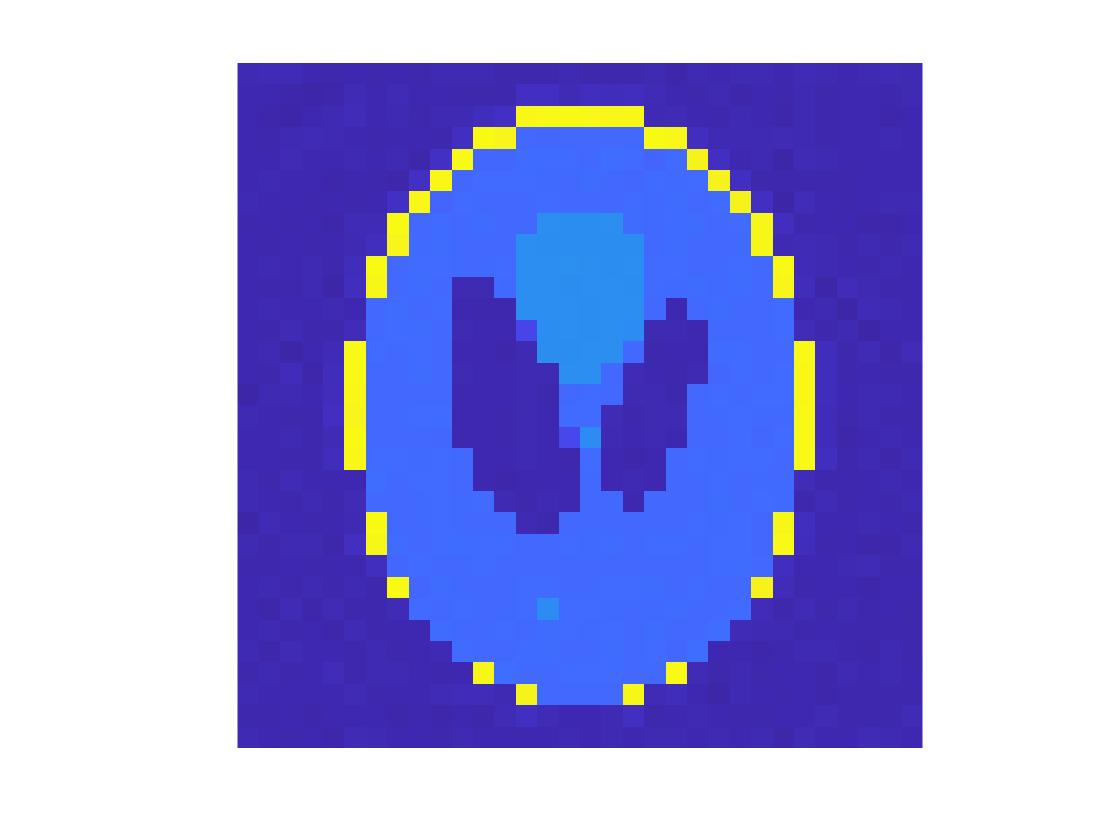}
    \includegraphics[width=0.33\textwidth]{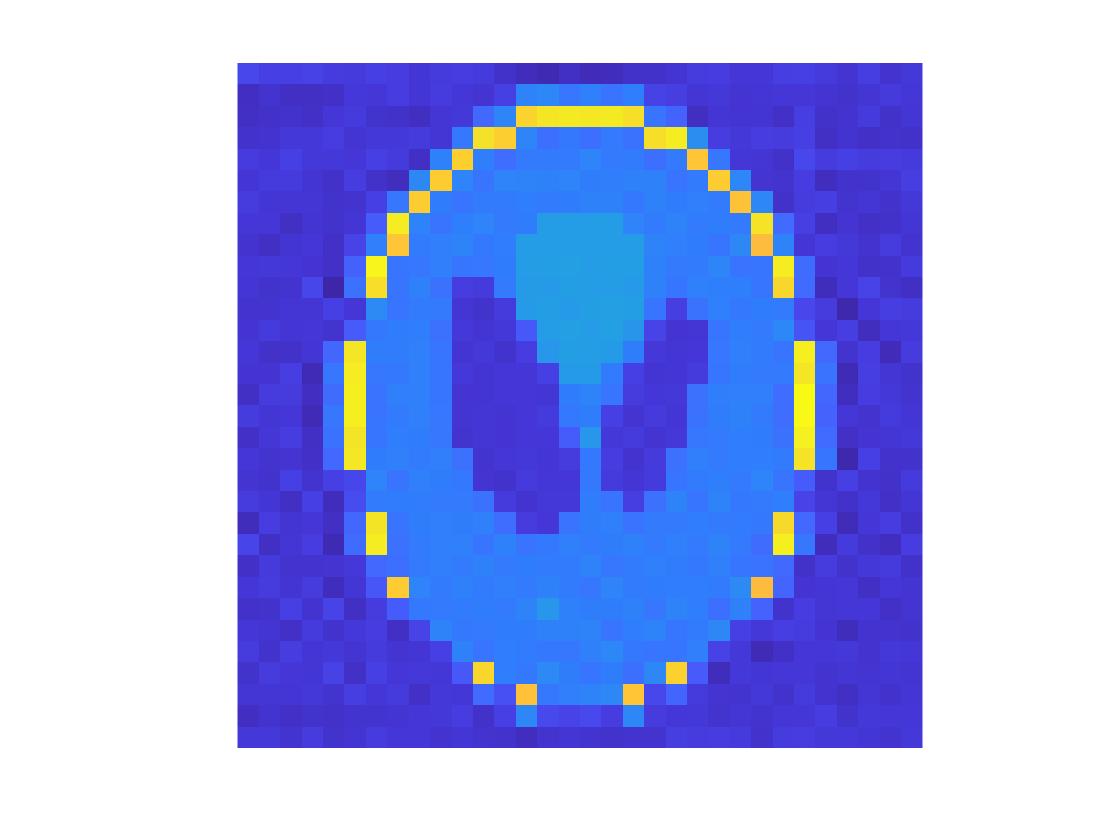}%
    \caption{Comparison of $32\times32$ Shepp-Logan phantom: true image (left), BCDS image (middle), BCD image (right).}
    \label{fig:BCDvsBCDS_im}
\end{figure}

Let BCD that exploits separability be called BCDS. In this section, we compare the running time of BCD and that of BCDS. In this subsection's numerical experiments, $budget=1000$ (used in \textbf{imfil} to put a limit on the number of function evaluations), and $N_A=10$. As we see in Figure \ref{fig:BCDvsBCDS_R}, BCDS dramatically speeds up the convergence because the separability allows us to solve a much smaller problem independently for one $r(i)$ at a time. The average running time of image reconstruction using BCD is $12.8s$, around the same as BCDS's $12.6s$. The time of geometry reconstruction using BCD is $1857.5s$, more then ten times longer than BCDS's $157.1s$. This is also the reason BCDS is discussed first in this section. In the remaining experiments, we always use the BCDS to reduce the running time. Moreover, the geometry errors and reconstruction errors of BCDS are both better than that of BCD. In Figure \ref{fig:BCDvsBCDS_im}, the phantom reconstructed by BCDS is much closer to the true image than the one from BCD. 

\par
A typical CT image using fan beam x-ray sources collects data at angles of one degree increment, from 0 to 359 degrees. In a perfect machine, the geometry parameters are precisely known each time the source is rotated to a new angle. In our experimental scenario, we assume the geometry parameters are only known approximately. To experiment with various scenarios, we assume that errors are introduced into the geometry parameters once every $\frac{360}{N_A}$ degrees. That is, with $N_A = 10$, errors occur once every 36 degrees. \par

Partially, $N_A$ depends on the precision of machine calibration. For a good machine, $N_A$ may be small. Also, $N_A$ may depend on how precise we measure the data. For example, two geometry parameters that are different in terms of double precision may be rounded to the same number in single precision. In this scenario, the difference could be small enough that we can treat the two geometry parameters as being equal. 

If the number of angles is larger than the true number of angles, we may end up solving a larger problem than we need. For example, if for every $36$ degrees only one geometry error is introduced, then $N_{A_{true}} = 10$. If we assume $N_A = 20$, the average of the two geometry errors per $36$ degrees would approximate the one true geometry error introduced in that set of angles. If the number of angles is smaller than the true number of angles, we end up solving a low dimension approximation. We do not seek to solve for the image exactly but aim to compute good approximations that yield much better results than not considering geometry parameters at all. In practice, we can consider the number of angles as a hyper-parameter that practitioners can set based on their expertise and knowledge of the machine calibration. In our problems, we assume to know the number of angles. \par
\subsection{Number of Angles} \label{sec:NA}
We compare the case where the number of angles is $5$, $10$, and $20$ respectively. We want to explore the differences in relative error of $r$, relative error of $x$, the convergence, and the image quality. \par
In this comparison, $budget=10$. The reason that we can use a such small budget is we are essentially solving one scalar of $r$ at a time. The recommended formula \cite{CTKelley} to set up a good guess for budget is $10*N^2$, where $N$ is the length of the solution vector. Since we solve for each component of the solution vector separately, the budget is just $10$ based on the recommended formula. Adjusting budget according to each particular problem as a hyper-parameter could further improve the quality of the result. Since we try to make apple-to-apple comparisons, we keep budget the same for all three cases.\par 
\begin{figure}{h}
    \centering
    \includegraphics[width=0.5\textwidth]{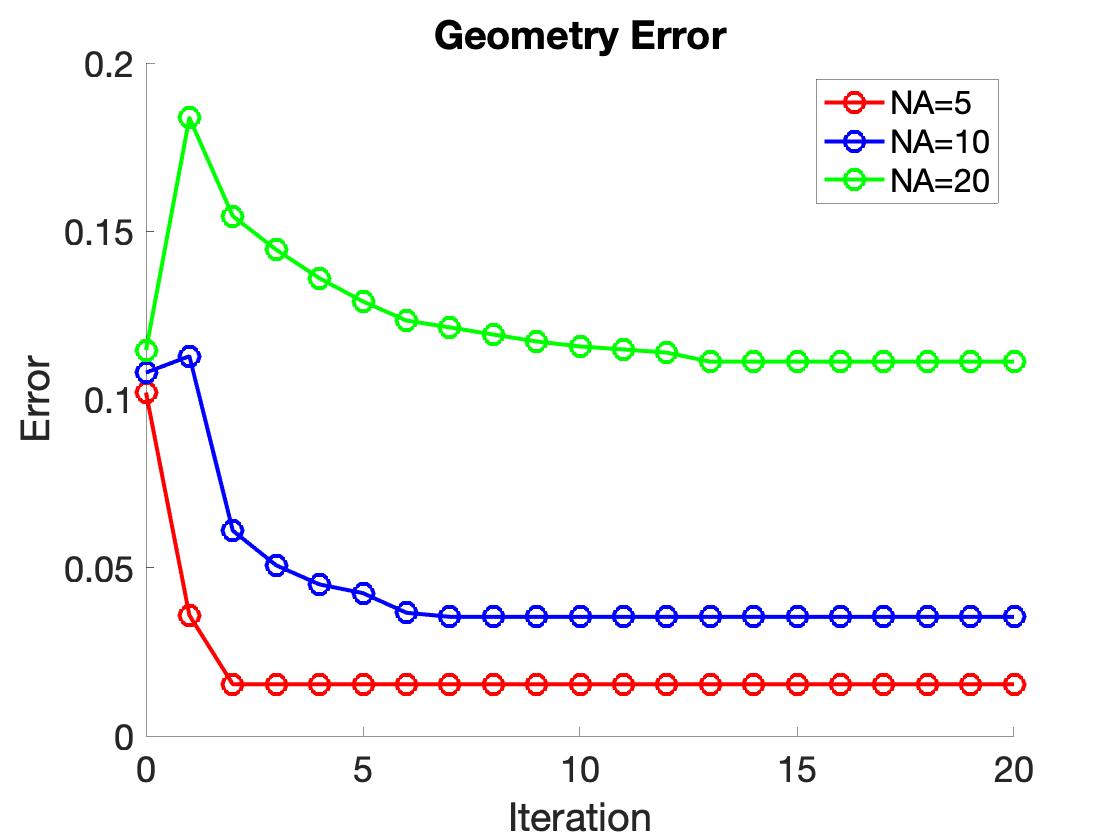}%
    \includegraphics[width=0.5\textwidth]{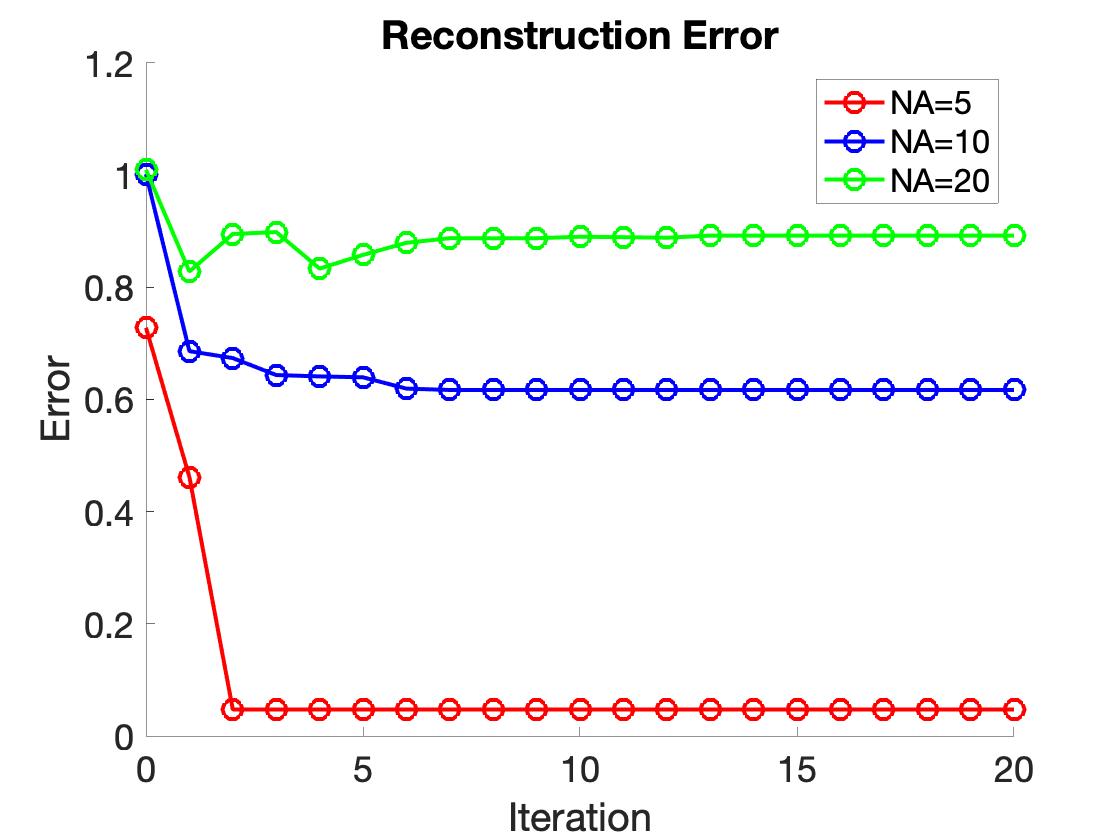}%
    \caption{Comparison of number of angles: geometry errors (left), reconstruction errors (right)}
    \label{fig:NA-R}
\end{figure}

Both geometry and image parameters converge for all three different number of angles, as shown in Figure \ref{fig:NA-R}. Although the errors when $N_A=10$ are not as low as when $N_A=5$, the reconstruction errors when $N_A=10$ still have a $40\%$ reduction comparing to its initial error level. When $N_A=20$, the geometry error has a huge spike at the first iteration. Although our algorithm has gradually decreased the geometry error after the first iteration, the resulting geometry error is almost the same as the geometry error of the initial guess. The reconstruction error drops from more than $100\%$ at the initial guess to below $85\%$, but the error climbed back up in the next few iterations. In the end, there was a $10\%$ drop in reconstruction errors comparing to the initial guess. There is one possible explanation for this unusual phenomenon when $N_A=20$. Let $r_k$ and $x_k$ represent the value of $r$ and $x$ at $k^{th}$ iteration in BCDS. First, given $x_0$, $r_1$ is calculated from the nonlinear least squares problem in Algorithm \ref{alg:AM}. $r_1$ is particularly bad because how bad $x_0$ is. Despite this $x_0$, BCDS is still improving the geometry parameter after the first iteration.

\subsection{Acceleration}
In this section, we compare BCDS, ABCDS, and BCDS with Anderson acceleration, where ABCDS represents the accelerated block coordinate descent that exploits separability of geometry parameters. The number of columns of the matrix $F_k$, also referred to as memory size in this paper, in Algorithm \ref{alg:AAunconstrained} is $m=5$. Since the linear least squares problem at Step $5$ of Algorithm \ref{alg:AAunconstrained} is relatively small, we directly use MATLAB backslash operator to solve it.  \par
There are two implementations of ABCDS that we compare. Then, we compare Anderson acceleration, ABCDS, and the BCDS method.

\subsubsection{ABCDS}
Since we can choose to apply the acceleration scheme on the image vector $x$ only or apply it on both geometry parameter $r$ and image vector $x$, we compare the two versions of ABCDS. We denote the former version of ABCDS as ABCDS-1. We denote the latter version that applies the acceleration on both image and geometry parameters as ABCDS-b. The experiment setup for this subsection is $budget = 100$ and $N_A=10$.

\begin{figure}[ht]
    \centering
    \includegraphics[width=0.5\textwidth]{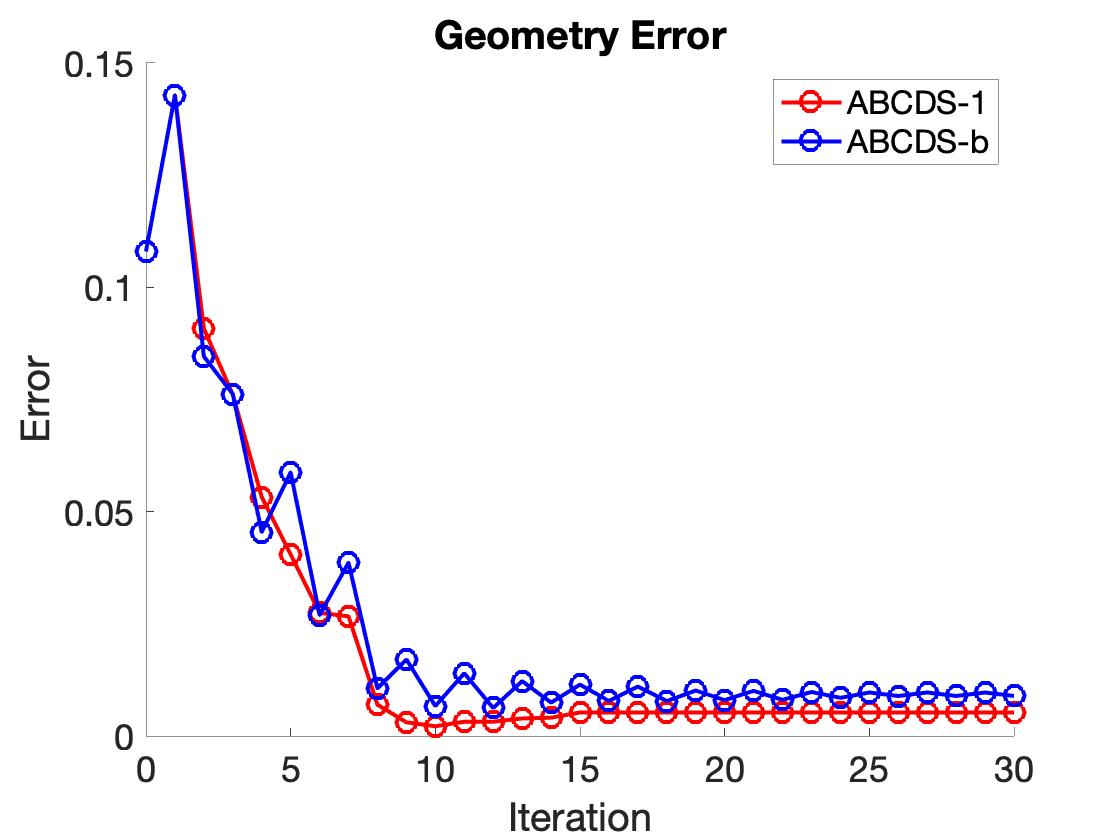}%
    \includegraphics[width=0.5\textwidth]{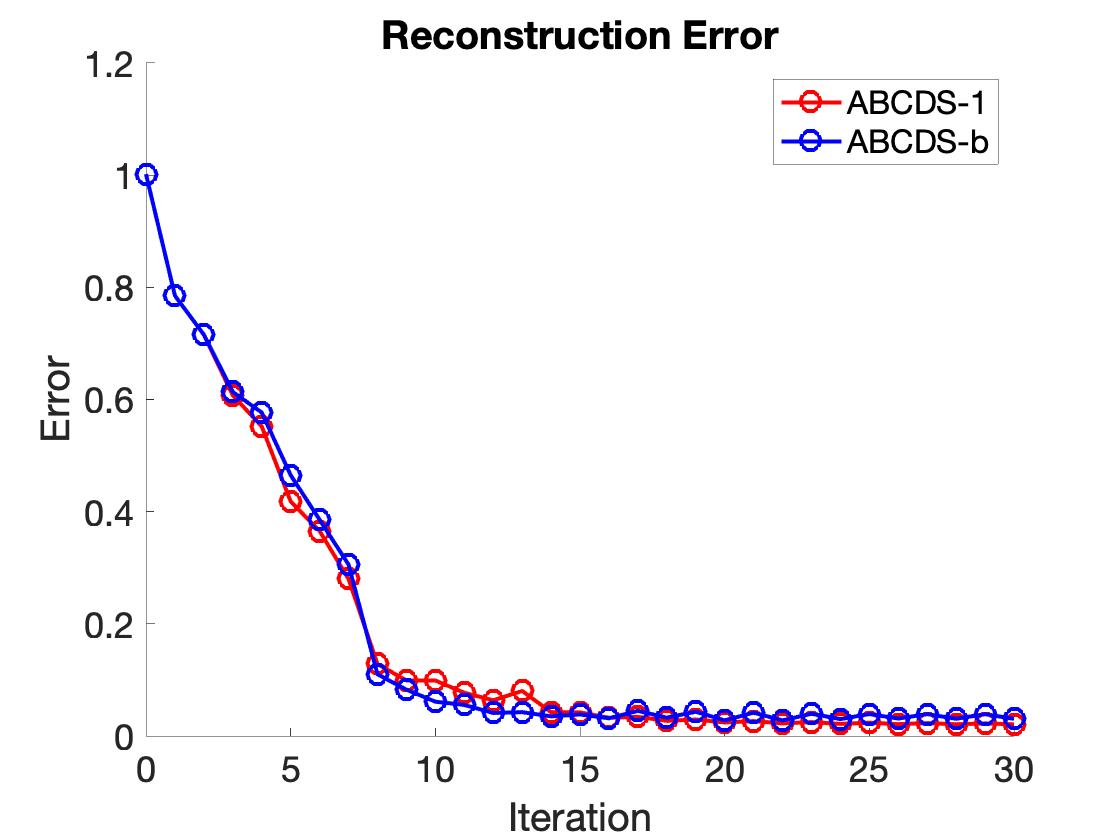}
    \caption{Comparison between ABCDS-1 and ABCDS-b: geometry errors (left), reconstruction errors (right) }
    \label{fig:imABCD-R}
\end{figure}

From Figure \ref{fig:imABCD-R}, we can tell that although both approaches converge and produce similar results at the end of iterations, the geometry errors of ABCDS-1 are slightly lower and more stable than that of ABCDS-b. Thus, applying the acceleration scheme on the image vector $x$ alone is enough to produce stable and  convergent results. For the sake of simplicity, when we mention ABCDS in the rest of the paper, it always refers to ABCDS-1 .  

\subsubsection{BCDS, ABCDS, and Anderson}
\begin{figure}[ht]
    \centering
    \includegraphics[width=0.5\textwidth]{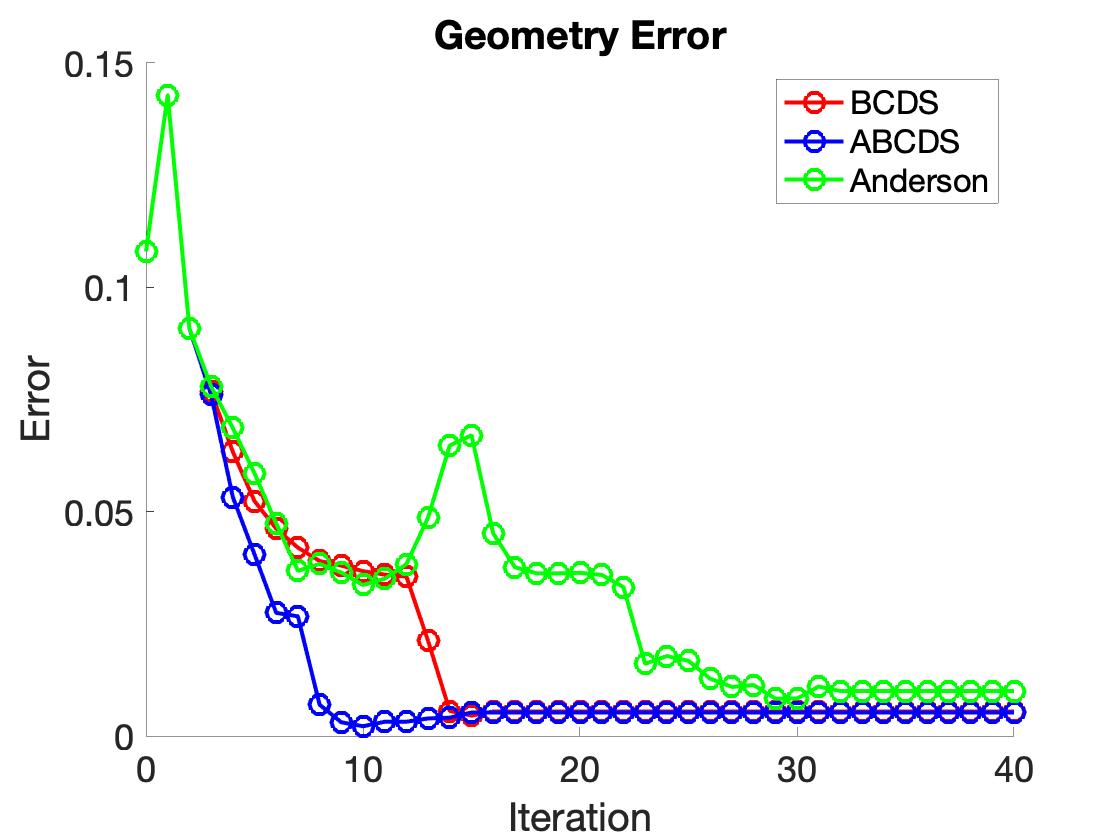}%
    \includegraphics[width=0.5\textwidth]{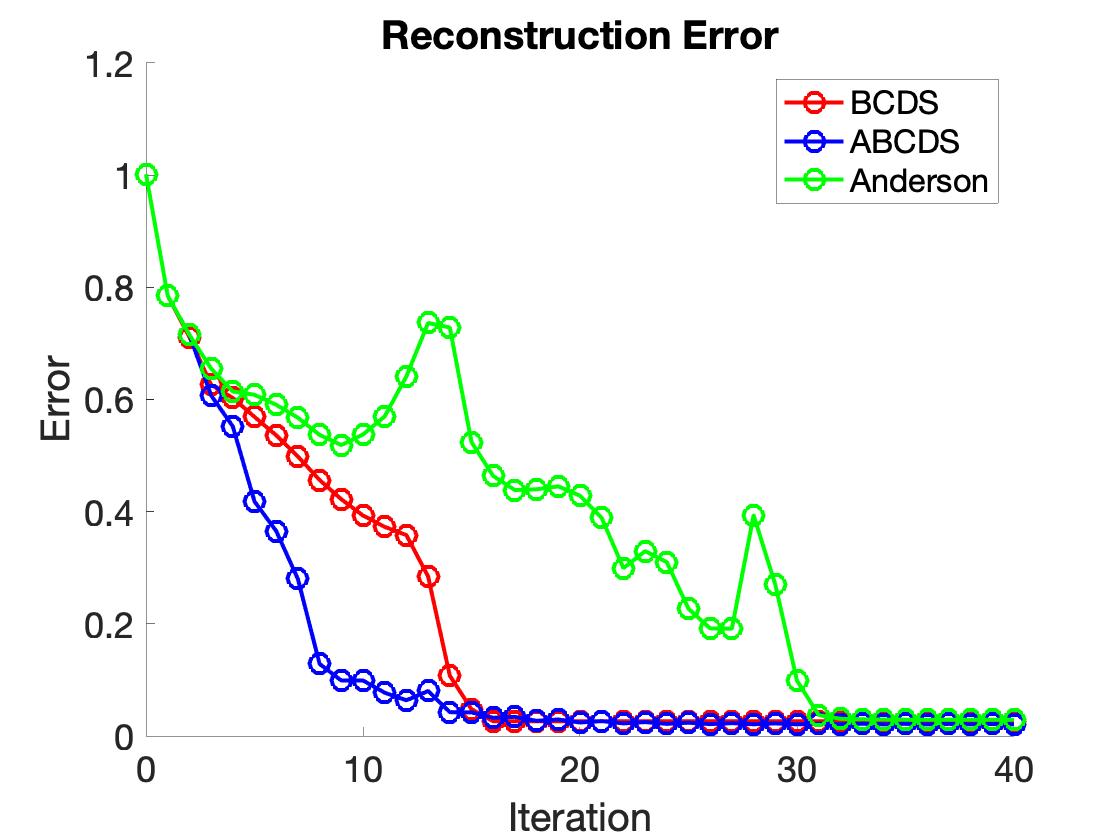}
    \caption{Comparison of BCDS, ABCDS, and Anderson: geometry errors (left), reconstruction errors (right)}
    \label{fig:BCDS-ABCD-R}
\end{figure}

In Figure \ref{fig:BCDS-ABCD-R}, geometry errors and reconstruction errors converge for all three methods. The geometry errors and the reconstruction errors of BCDS converge around $15^{th}$ iteration. The both geometry and reconstruction errors of ABCDS converge around $10^{th}$ iteration. Anderson acceleration converges slowest because there are some spikes around $15^{th}$ iteration for both geometry and reconstruction errors. There are safeguards that can ensure global convergence of the type I Anderson acceleration \cite{boyd}. However, in our problem, Anderson acceleration does converge but does not accelerate the convergence. This could be an interesting challenge to be explored in the future. For our problem, ABCDS is a better acceleration scheme its convergence and acceleration effects.

\subsection{Regularization}
\begin{figure}[ht]
    \centering
    \includegraphics[width=0.5\textwidth]{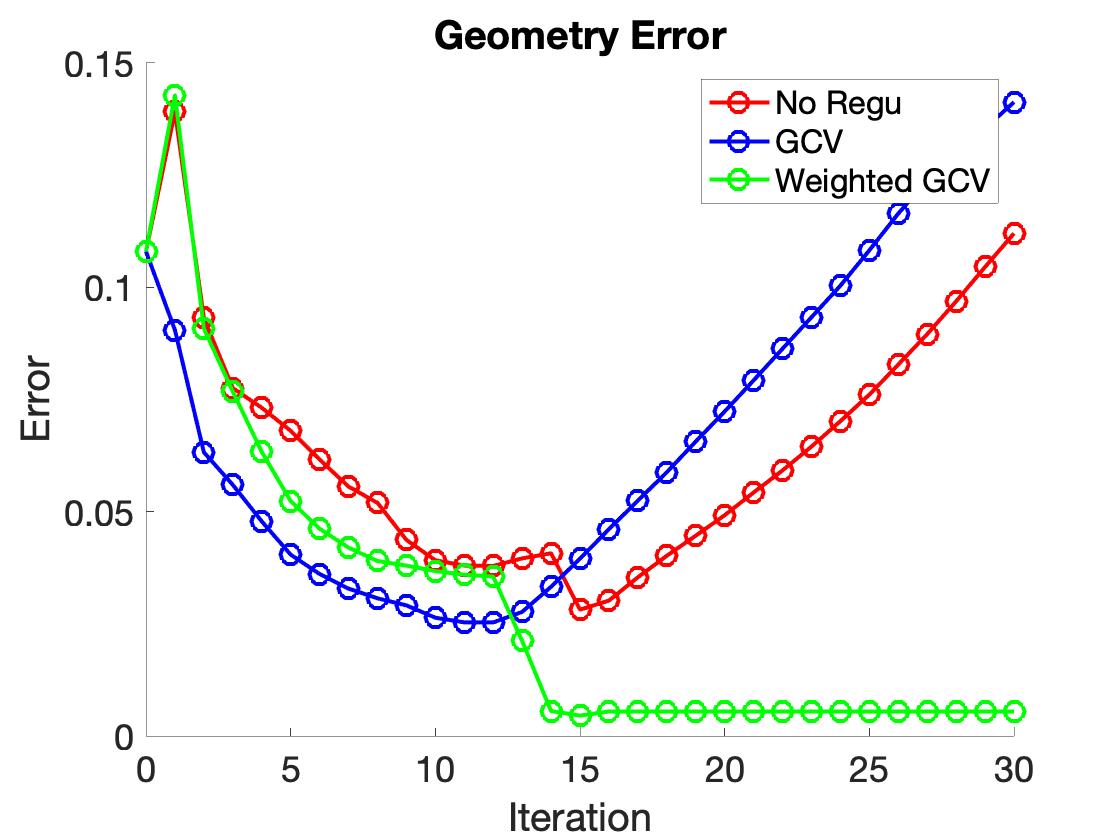}%
    \includegraphics[width=0.5\textwidth]{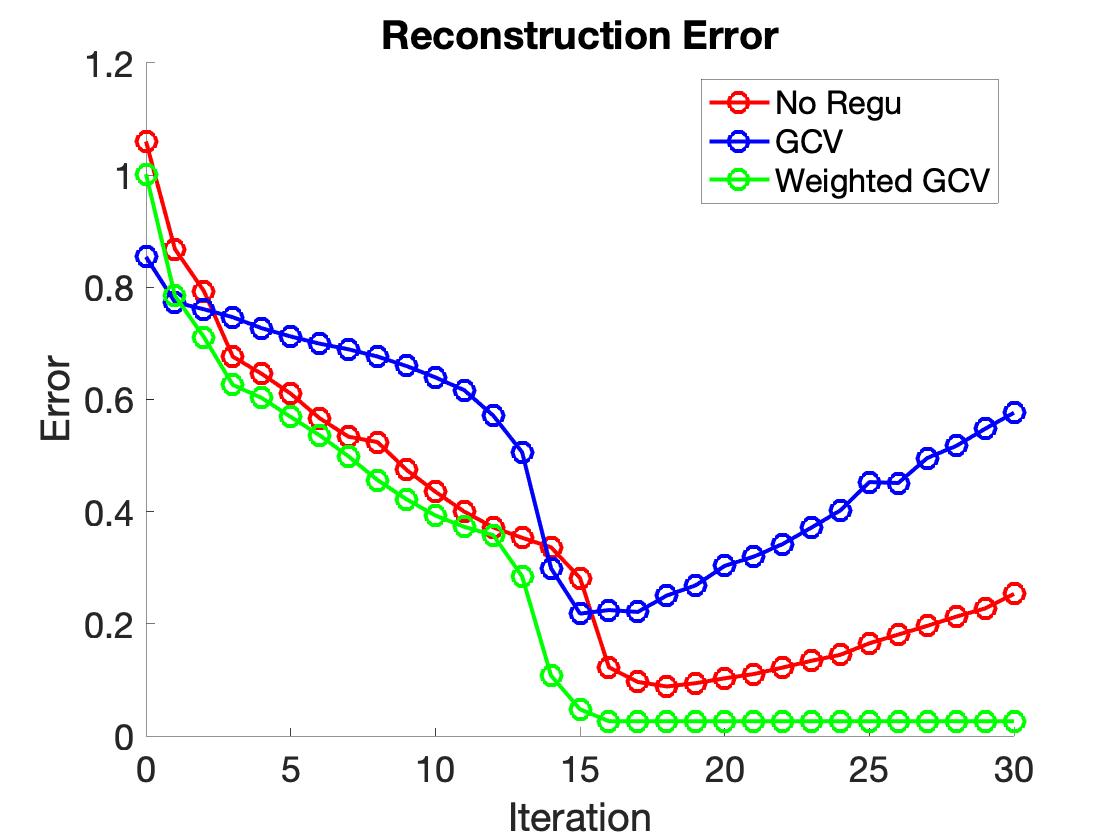} 
    \caption{Comparison of regularization: geometry errors (left), reconstruction errors (right)  }
    \label{fig:Reg-R}
\end{figure}

In this section, we compare BCDS with different regularization parameters: no regularization, GCV, and  weighted-GCV, as shown in Figure \ref{fig:Reg-R}. The reason we use BCDS without the acceleration methods is that we want to see the direct impact of regularization on the alternating minimization scheme. In this experiment, $N_A=10$ and $budget=100$. \par

When there is no regularization, \textbf{IRhybrid\_lsqr} is essentially an LSQR algorithm, which exhibits semi-convergence behavior. When GCV is used instead of the weighted-GCV, the geometry and image parameter errors also exhibit semi-convergence behavior. The weighted-GCV helps stabilize LSQR's convergence and thus produces the best result. 
\subsection{Imfil Budget}
We explore the effect of evaluation budgets in the nonlinear least squares solver on the geometry and reconstruction errors. We set $N_A=10$, and use $budget=10,100,1000,10000$. Since the budget size may greatly affect the nonlinear least squares solutions, we explore its effects on BCDS without any acceleration. \par

\begin{figure}[ht]
    \centering
    \includegraphics[width=0.5\textwidth]{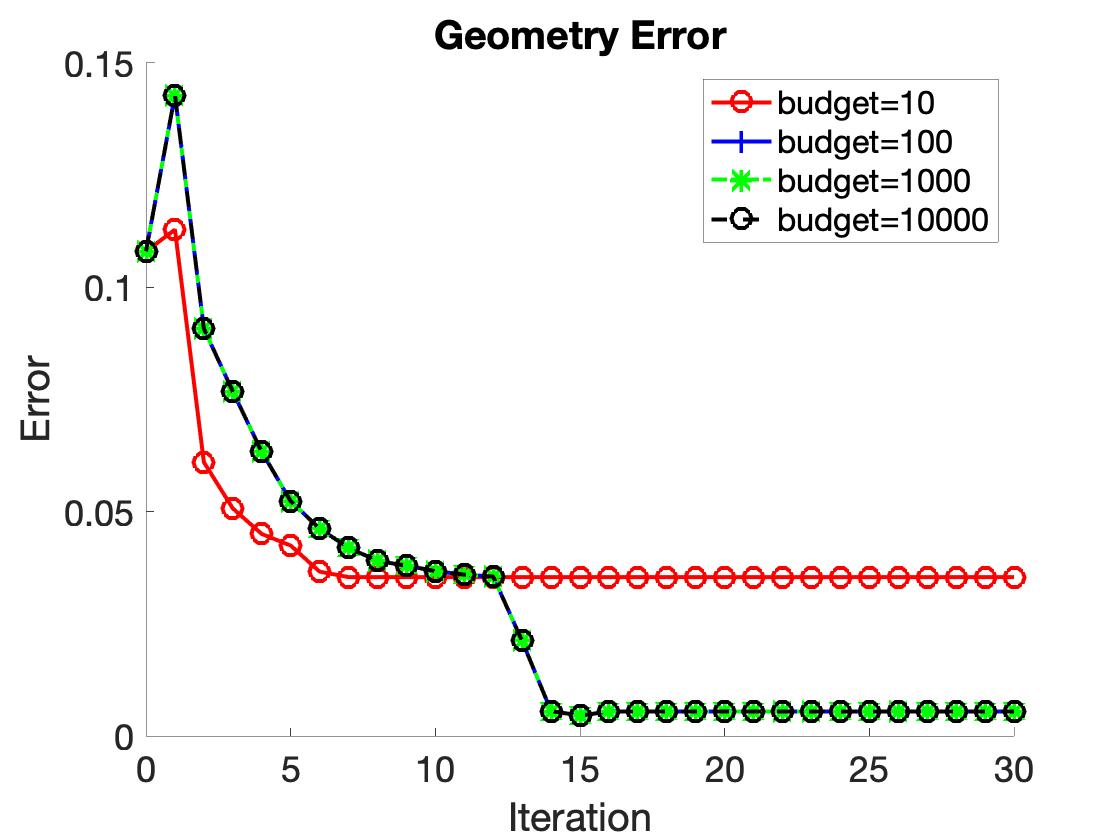}%
    \includegraphics[width=0.5\textwidth]{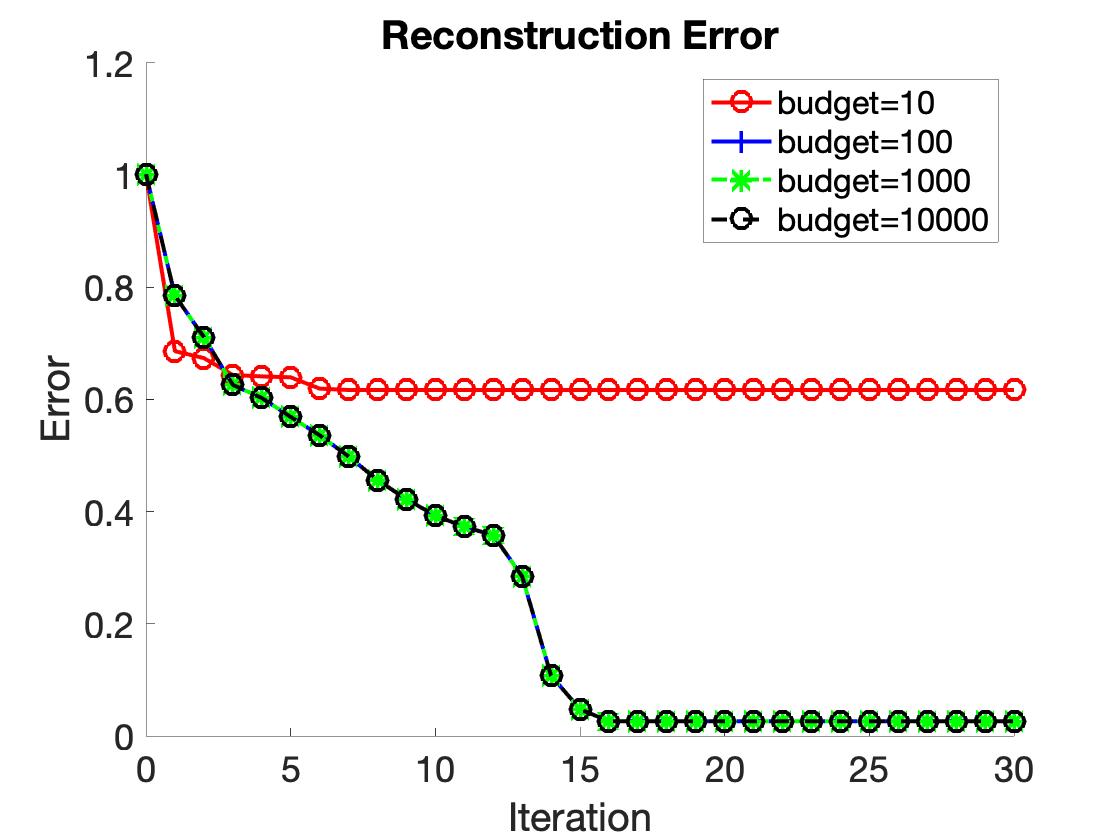} 
    \caption{Comparison of budgets: geometry errors (left), reconstruction errors (right)}
    \label{fig:budget-R}
\end{figure}

As we can see in Figure \ref{fig:budget-R}, $100$, $1000$, and $10000$ are equally good. Both geometry and reconstruction errors are very small when $100$, $1000$, $10000$ are used. Thus, $100$ is the best budget out of the four because any more evaluation beyond $100$ does not make the solution better. When budget is $1000$ and $10000$ respectively, we wasted many evaluations without making any progress.  When $10$ is used, even though we get convergent results earlier, the small budget causes the algorithm to terminate before it finds a better solution. 

\subsection{Nonlinear Least Squares Solver}
In this section, we compare BCDS with different nonlinear least squares solvers – \textbf{fminbnd} and \textbf{imfil}. Set $N_A=10$ and $budget = 100$.
\begin{figure}[ht]
    \centering
    \includegraphics[width=0.49\textwidth]{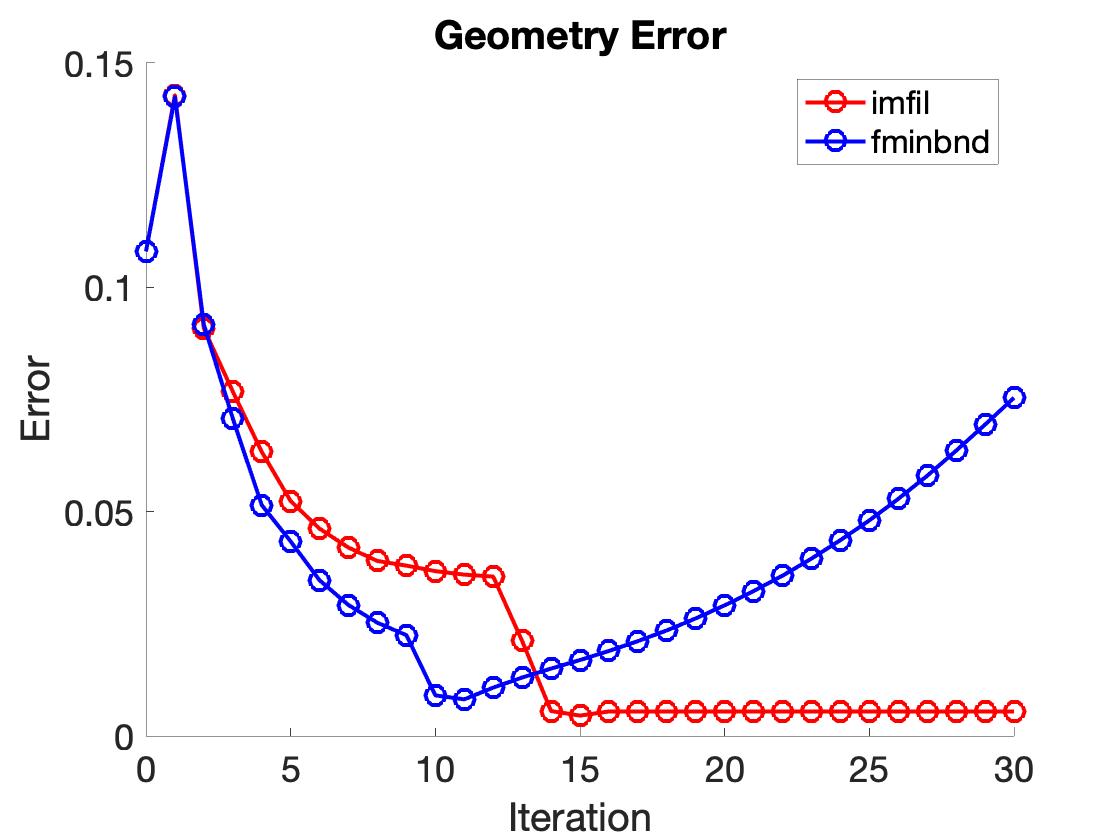} %
    \includegraphics[width=0.49\textwidth]{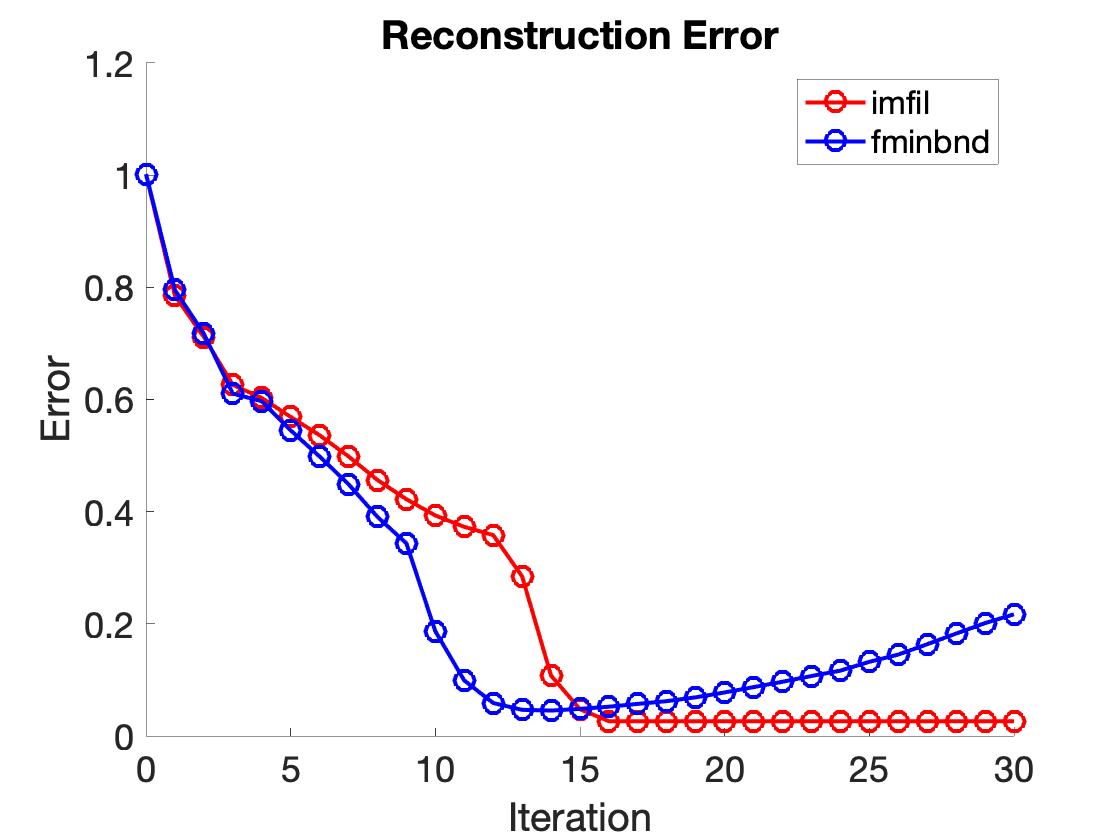} 
    \caption{Comparison of nonlinear least squares solvers: geometry errors (left), reconstruction errors (right)}
    \label{fig:fmin-im-R}
\end{figure}

In Figure \ref{fig:fmin-im-R}, \textbf{fminbnd} exhibits semi-convergence behavior whereas \textbf{imfil} converges. Under our problem settings, \textbf{fminbnd} reconstructs geometry parameters around $10$s faster than \textbf{imfil} on average. However, \textbf{fminbnd} tends to exhibit semi-convergence behavior whereas iterations in \textbf{imfil} are more stable. Since in practical cases it is impossible to know the exact iteration, at which the geometry errors are lowest in the semi-convergence case, \textbf{imfil} is more reliable in solving the nonlinear least squares problem. 

\subsection{Geometry Parameters: $d$ and $\delta\theta$}
\begin{figure}[ht]
    \centering
    \includegraphics[width=0.3\textwidth]{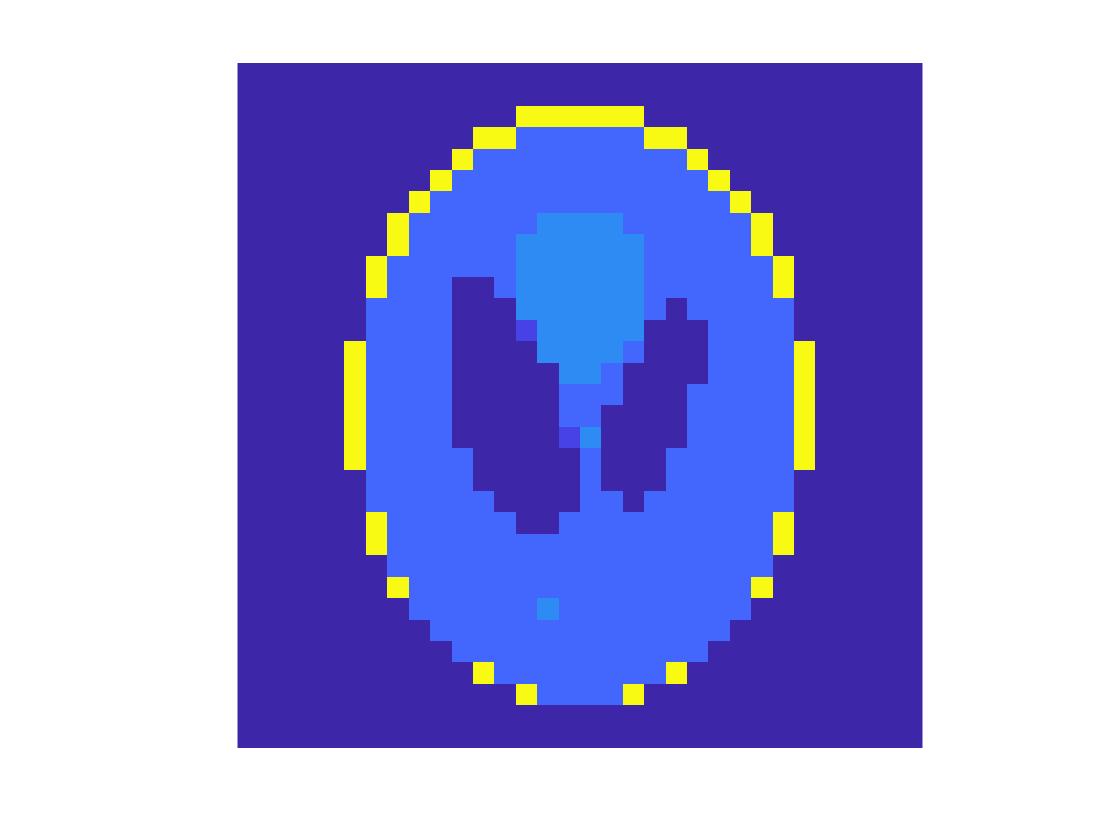}%
    \includegraphics[width=0.3\textwidth]{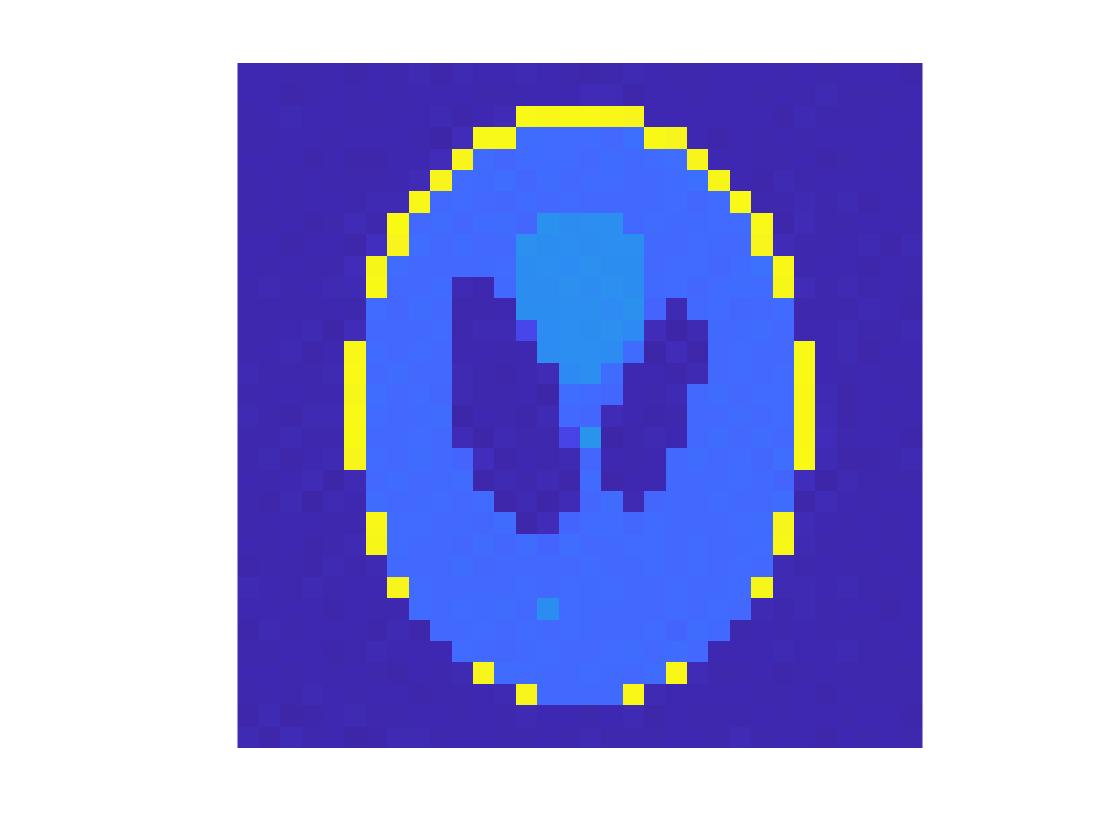}
    \includegraphics[width=0.3\textwidth]{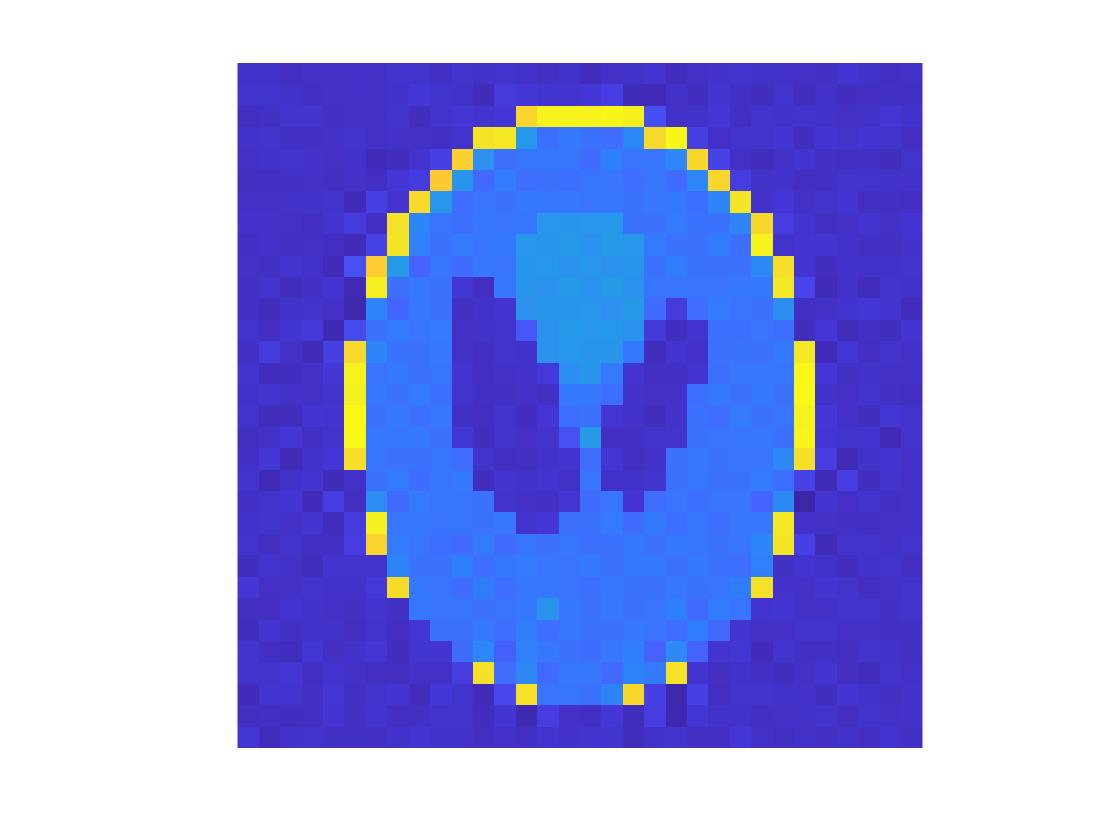}%
    \includegraphics[width=0.3\textwidth]{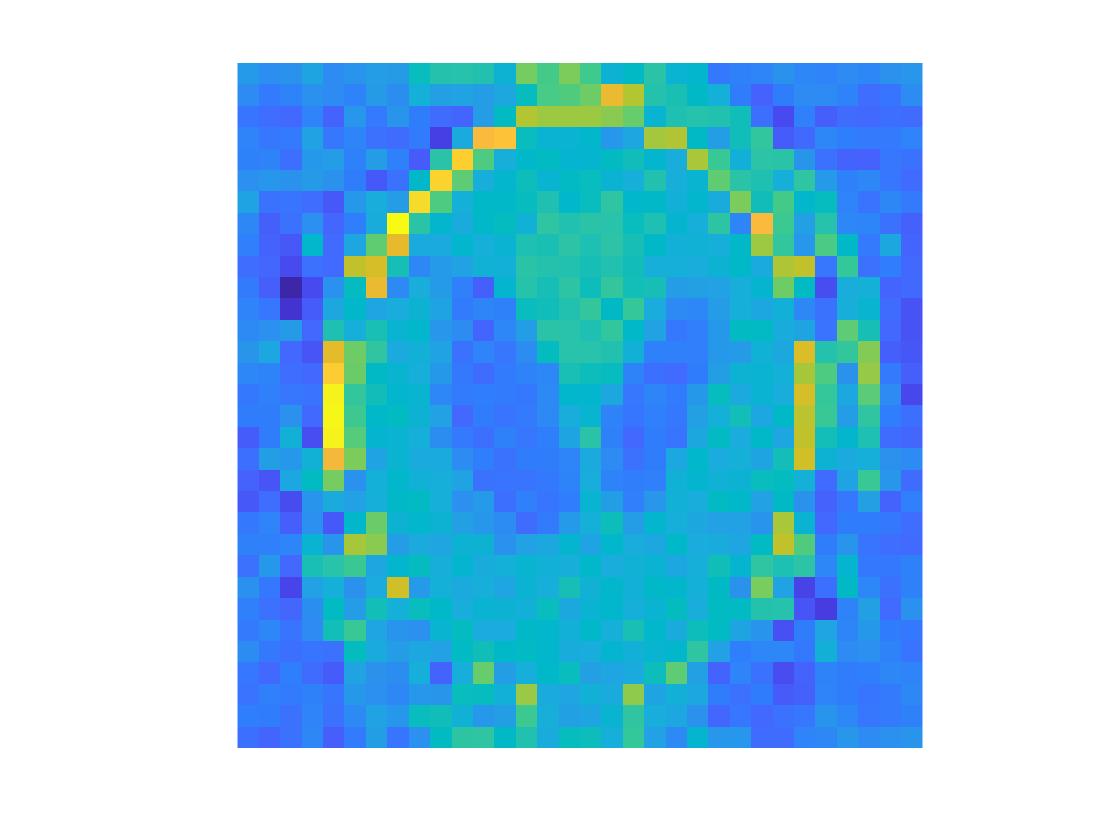} 
    \caption{True image (top left), image obtained by considering both $\delta\theta$ and $d$ (top right), image obtained by using correct $d$ but incorrect $\delta\theta$ (bottom left), image obtained by using incorrect geometry parameters $d$ and $\delta\theta$ (bottom right).}
    \label{fig:Rtheta_image}
\end{figure}

We show that our alternating scheme as well as the accelerated version can also be applied to the case when geometry parameters $r = (d,\texorpdfstring{\delta}{del}\texorpdfstring{\theta}{theta})$.The nonlinear least squares problem becomes: $$\underset{x,\delta\theta, d}{\arg\min}||A(\delta\theta,d)x-b||. $$
In all previous experiments we assume there was no error in angles. If the angles contain errors as well, the image quality could also be affected, as illustrated in Figure \ref{fig:Rtheta_image}. The experiment setting is $N_A=10$ and $budget = 100$.

\begin{figure}[ht]
    \centering
    \includegraphics[width=0.33\textwidth]{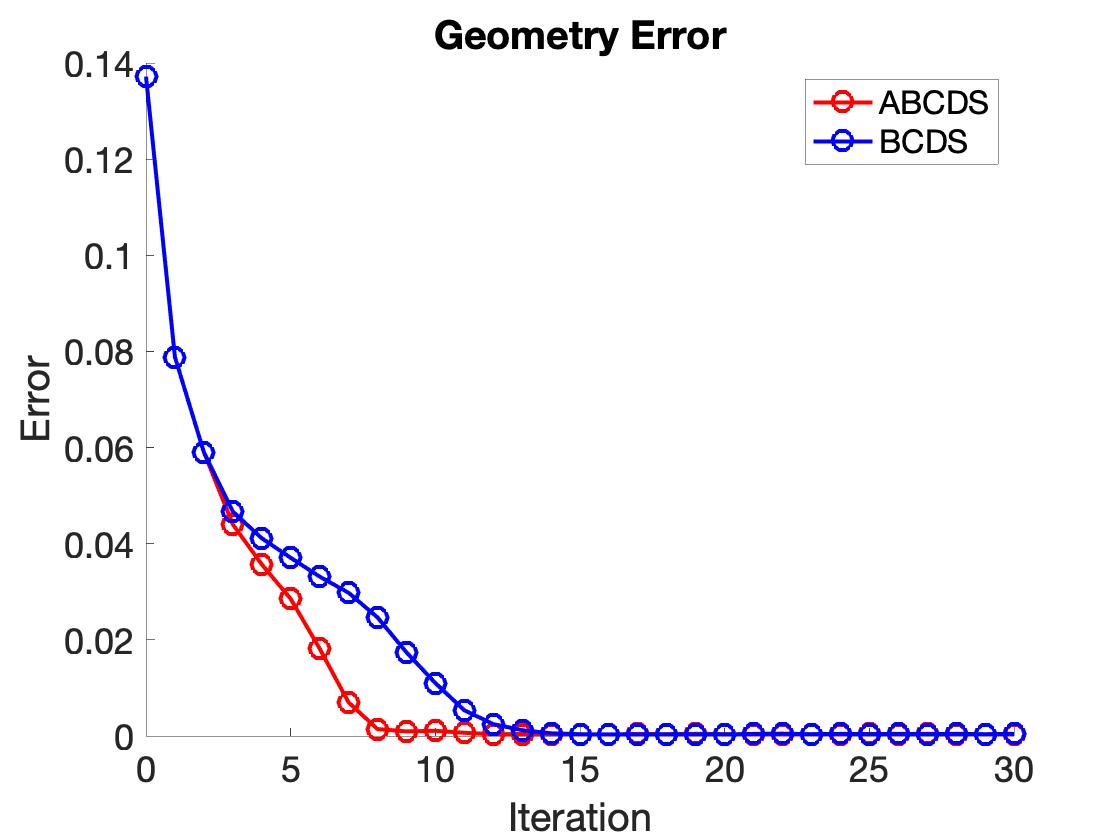}%
    \includegraphics[width=0.33\textwidth]{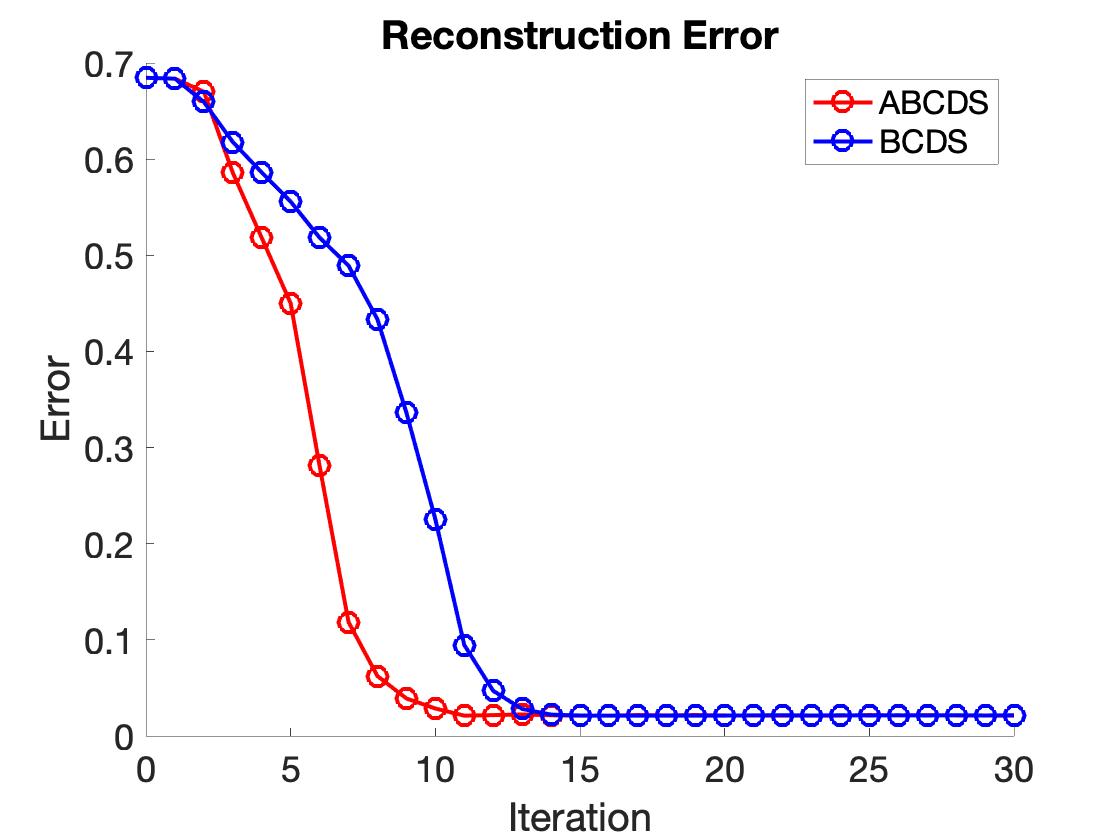}%
    \includegraphics[width=0.33\textwidth]{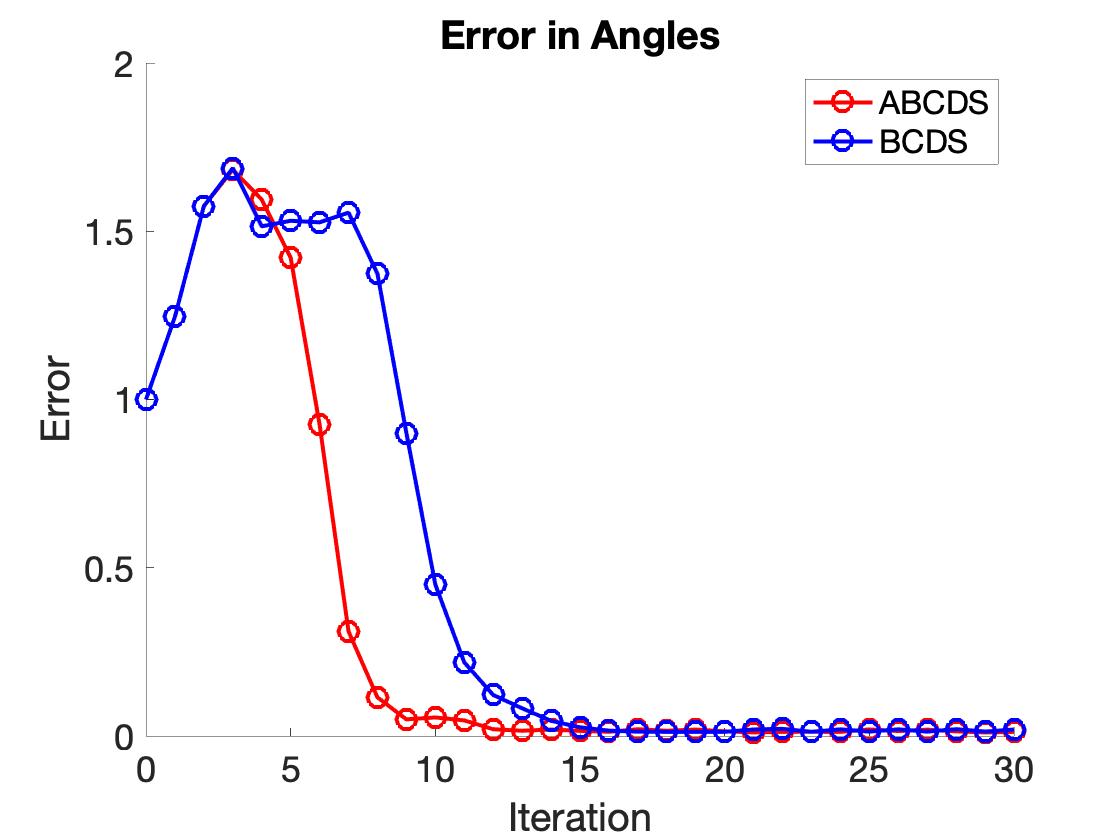}%
    \caption{Comparison of BCDS and ABCDS when correcting for errors in source-to-object distance and angles: geometry errors (left), reconstruction errors (middle), errors in angles (right)}
    \label{fig:Rtheta_errors}
\end{figure}

As shown in Figure \ref{fig:Rtheta_errors}, both ABCDS and BCDS converge for all three parameters $d$, $\delta\theta$, and $x$. ABCDS has accelerated the convergence for all three geometry parameters, especially for the reconstruction errors. The only difference between this problem, $r=(d,\delta\theta)$, and the previous problem, $r=d$, is we solve for two geometry parameters for each nonlinear least squares problem instead of one. 

\section{Conclusion} \label{sec:conclusions}
In this paper, we propose an accelerated alternating minimization scheme to solve X-ray tomography problems. The linear least squares problem is solved by a weighted hybrid LSQR algorithm with Tikhonov regularization. The nonlinear least squares problem is solved by implicit filtering. We also investigated ABCDS and Anderson mixing to accelerate convergence. We have the following findings:
\begin{itemize}
    \item[1.] BCDS runs faster than the normal BCD because the separability of parameters allow us to solve each entry of $r$ independently. Also, since solving for each parameter separately makes the nonlinear least squares problem much easier than solving all parameters at once, BCDS is faster at finding a good solution and converge faster than BCD.
    \item[2.] ABCDS is much less computationally costly than BCDS with Anderson acceleration because Anderson acceleration requires solving a linear least squares problem in each iteration when choosing the weight. Although both methods converge, ABCDS have better acceleration effects than Anderson acceleration.
    \item[3.] The weighted hybrid LSQR algorithm with Tikhonov regularization stabilizes the convergence more than applying an unweighted GCV to the LSQR algorithm. When no regularization method is used, the LSQR algorithm exhibits semi-convergence behavior. It reconstructs the solution at earlier iterations but noise at later iterations.
    \item[4.] Choosing an appropriate budget for implicit filtering is important. When the budget is chosen too small, better solutions are not explored. When budget is chosen too big, many function evaluations are wasted without making progress. A suggested number in the \textbf{imfil} documentation is $10*N^2$, where $N$ is the length of $r$. We found this formula does not always give the appropriate budget. Since we solve for each geometry parameter using separability, the dimension of each small problem is one. But, we have found $100$ to be the best budget for $32\times32$ test problem with $N_A=10$.
    \item[5.] \textbf{imfil} is a better nonlinear least squares solver for our algorithm than MATLAB's \textbf{fminbnd} because the latter tends to exhibit semi-convergence behaviors. 
\end{itemize}

The ABCDS method has shown its success in our tomographic reconstruction problems. We believe this algorithm can be used to effectively solve X-ray tomography problems that have variations in the geometry parameter. A future direction towards more improvement would be adapting, applying, and advancing this algorithm on X-ray images produced in clinical trials.

\section*{Acknowledgments}
This work was done as part of an undergraduate honors thesis project at Emory University, and was partially supported by the U.S. National Science Foundation under Grant DMS-1819042. The author would like to thank Chang Meng and James Nagy for their guidance and mentorship.
\bibliographystyle{siamplain}
\bibliography{references.bib}
\end{document}